\documentclass[pdflatex,sn-mathphys-num]{sn-jnl}

\usepackage{graphicx}%
\usepackage{multirow}%
\usepackage{amsmath,amssymb,amsfonts}%
\usepackage{amsthm}%
\usepackage{mathrsfs}%
\usepackage[title]{appendix}%
\usepackage{textcomp}%
\usepackage{manyfoot}%
\usepackage{algpseudocode}%
\usepackage{listings}%
\usepackage{hhline}

\usepackage{anyfontsize} 

\usepackage{adjustbox}           
\usepackage{booktabs}

\usepackage[T1]{fontenc}
\usepackage{type1ec}
\usepackage{graphicx} 
\usepackage{url}
\usepackage{rotating}
\usepackage{tabularx}
\usepackage{amsmath}
\usepackage{amsfonts}
\usepackage{adjustbox}
\usepackage{multirow}
\usepackage{hyperref}
\usepackage{etoolbox}
\usepackage{longtable}

\usepackage[g]{esvect} 

    \usepackage{xfrac}

\usepackage{tabularx}
\newcolumntype{Y}{>{\centering\arraybackslash}X}

\usepackage[linesnumbered,ruled,
			vlined,
			nofillcomment]{algorithm2e} 
\SetKw{KwAnd}{and}
\SetKwHangingKw{Let}{let}
\SetArgSty{textnormal}
\DontPrintSemicolon

\usepackage{afterpage}
\usepackage{lscape}

\usepackage{mathtools} 

\allowdisplaybreaks 

\usepackage{listings} 
\usepackage{textcomp} 
\usepackage{bera}
\usepackage{listings}
\usepackage[dvipsnames]{xcolor}

\usepackage{enumitem} 

\colorlet{punct}{red!60!black}
\definecolor{background}{HTML}{EEEEEE}
\definecolor{delim}{RGB}{20,105,176}
\colorlet{numb}{magenta!60!black}

\lstdefinelanguage{json}{
    basicstyle=\normalfont\ttfamily,
    literate=
     *{0}{{{\color{numb}0}}}{1}
      {1}{{{\color{numb}1}}}{1}
      {2}{{{\color{numb}2}}}{1}
      {3}{{{\color{numb}3}}}{1}
      {4}{{{\color{numb}4}}}{1}
      {5}{{{\color{numb}5}}}{1}
      {6}{{{\color{numb}6}}}{1}
      {7}{{{\color{numb}7}}}{1}
      {8}{{{\color{numb}8}}}{1}
      {9}{{{\color{numb}9}}}{1}
      {:}{{{\color{punct}{:}}}}{1}
      {,}{{{\color{punct}{,}}}}{1}
      {\{}{{{\color{delim}{\{}}}}{1}
      {\}}{{{\color{delim}{\}}}}}{1}
      {[}{{{\color{delim}{[}}}}{1}
      {]}{{{\color{delim}{]}}}}{1},
}

\usepackage{rotating} 

\makeatletter
\newcommand{\subalign}[1]{%
	\vcenter{%
		\Let@ \restore@math@cr \default@tag
		\baselineskip\fontdimen10 \scriptfont\tw@
		\advance\baselineskip\fontdimen12 \scriptfont\tw@
		\lineskip\thr@@\fontdimen8 \scriptfont\thr@@
		\lineskiplimit\lineskip
		\ialign{\hfil$\m@th\scriptstyle##$&$\m@th\scriptstyle{}##$\hfil\crcr
			#1\crcr
		}%
	}%
}
\makeatother

\usepackage{array}
\newcolumntype{L}[1]{>{\raggedright\let\newline\\\arraybackslash\hspace{0pt}}m{#1}}
\newcolumntype{C}[1]{>{\centering\let\newline\\\arraybackslash\hspace{0pt}}m{#1}}
\newcolumntype{R}[1]{>{\raggedleft\let\newline\\\arraybackslash\hspace{0pt}}m{#1}}

\usepackage{bera}
\usepackage{listings}
\usepackage[dvipsnames]{xcolor}

\usepackage{amsthm}
\newtheoremstyle{noind}
{5pt}
{5pt}
{\itshape}
{}
{\bfseries}
{.}
{2pt} 
{}

\newtheoremstyle{plain_ind}
{5pt}
{5pt}
{}
{15pt}
{\bfseries}
{.}
{2pt} 
{}

\theoremstyle{noind}

\newtheorem{example}{{Example}}

\theoremstyle{plain_ind}

    \usepackage{thmtools, thm-restate}

\usepackage{tikz}
\usetikzlibrary{positioning}
\tikzset{main node/.style={text centered},}

\newcommand*\circled[1]{\tikz[baseline=(char.base)]{
            \node[shape=circle,draw,inner sep=2pt] (char) {#1};}}
\usetikzlibrary{shapes.geometric} 

\makeatletter
\DeclareRobustCommand
  \rvdots{\vbox{\baselineskip4\p@ \lineskiplimit\z@
    \hbox{.}\hbox{.}\hbox{.}}}
\makeatother

\newcounter{tuplesno}

\newif\ifReview
\Reviewtrue 
\usepackage[normalem]{ulem}
\usepackage{fancybox}

\newcommand{\red}[1]{\textcolor{red}{#1}}

\ifReview
    \newcommand{\mycomment}[2]{\newline\noindent\shadowbox{\begin{minipage}{\dimexpr0.98\textwidth-\shadowsize-2\fboxrule-2\fboxsep}\textbf{#1: #2}\end{minipage}}}

    \newcommand{\rem}[1]{\red{\sout{#1}}}
\else
    \newcommand{\mycomment}[2]{}

    \newcommand{\tirar}[1]{}
    \newcommand{\rem}[1]{}
\fi

\newcommand{\metodo}{Lathe}

\begin{document}

\title[Article Title]{Transformer-based Ranking Approaches for Keyword Queries over Relational Databases}


\author*[]{\fnm{Paulo} \sur{Martins}}\email{paulo.martins@icomp.ufam.edu.br}
\author[]{\fnm{Altigran} \sur{da Silva}}\email{alti@icomp.ufam.edu.br}
\author[]{\fnm{Johny} \sur{Moreira}}\email{johny.moreira@icomp.ufam.edu.br}
\author[]{\fnm{Edleno} \sur{de Moura}}\email{edleno@icomp.ufam.edu.br}


\affil[]{\orgdiv{Institute of Computing}, \orgname{Federal University of Amazonas}, \orgaddress{\street{Street}, \city{Manaus}, \postcode{100190}, \state{Amazonas}, \country{Brazil}}}


\abstract{Relational Keyword Search (R-KwS) systems enable naive/informal users to explore and retrieve information from relational databases without requiring schema knowledge or query-language proficiency. Although numerous R-KwS methods have been proposed, most still focus on queries referring only to attribute values or primarily address performance enhancements, providing limited support for queries referencing schema elements. We previously introduced \textit{Lathe}, a system that accommodates schema-based keyword queries and employs an \textit{eager CJN evaluation} strategy to filter out spurious Candidate Joining Networks (CJNs). However, Lathe still faces challenges in accurately ranking CJNs when queries are ambiguous. In this work, we propose a new \textit{transformer-based ranking} approach that provides a more context-aware evaluation of Query Matches (QMs) and CJNs. Our solution introduces a \textit{linearization} process to convert relational structures into textual sequences suitable for transformer models. It also includes a \textit{data augmentation} strategy aimed at handling diverse and ambiguous queries more effectively. Experimental results, comparing our transformer-based ranking to Lathe’s original Bayesian-based method, show significant improvements in recall and $R@k$, demonstrating the effectiveness of our neural approach in delivering the most relevant query results.}

\maketitle

\section{Introduction}\label{chap:intro}

\textit{Keyword Search over Relational Databases} (R-KwS) enables naive/informal users to retrieve information from relational databases (DBs) without any knowledge about schema details or query languages. The success of search engines shows that untrained users are at ease using keyword search to find information of interest. 
However, enabling users to search relational DBs using keyword queries is a challenging task, because the information sought frequently spans multiple relations and attributes, depending on the schema design of the underlying DB. As a result, R-KwS systems must determine which pieces of information to retrieve from the database. They must also figure out how to connect these pieces to provide a relevant answer to the user.

Over the last two decades, the R-KwS was extensively studied in academia, which lead to several improvements in performance and effectiveness. A well-known approach for R-KwS is to generate Candidate Joining Networks (CJNs), which are networks of joined database relations that are translated into SQL queries whose results provide an answer to the input keyword query. The first algorithm for CJN generation is CNGen, which was first presented in the DISCOVER system \cite{Hristidis@VLDB02DISCOVER}, but was later adopted by most R-KwS systems \cite{Agrawal@ICDE02DBXplorer,Hristidis@VLDB03Efficient,Luo@SIGMOD07Spark,Coffman@KEYS10CD}. Despite the possible large number of CJNs, most works in the literature focused on improving CJN evaluation and ranking the returned results from the database, which can be seen as Joining Networks of Tuples (JNTs), instead. Specifically, DISCOVER II \cite{Hristidis@VLDB03Efficient}, SPARK \cite{Luo@SIGMOD07Spark}, and CD \cite{Coffman@KEYS10CD} used information retrieval (IR) style score functions to rank the top-K JNTs. KwS-F \cite{Baid@VLDB10KwSF} imposed a time limit for the evaluation of CJNs, returning potentially partial results, as well as a summary of CJNs that have yet to be evaluated.
Later, CNRank~\cite{Oliveira@ICDE15CNRank} introduced a CJN ranking, requiring only the top-ranked CJNs to be evaluated. 
MatCNGen~\cite{Oliveira@ICDE18MatCNGen,Oliveira@TKDE20} proposed a novel method for generating CJNs that efficiently enumerated the possible matches for the query in the DB. These \textit{Query Matches} (QMs) are then used to guide the CJN generation process, greatly decreasing the number of CJNs generated and improving the performance of the CJN evaluation.

Building on these foundations, we previously proposed \textit{Lathe}~\cite{Martins@ICDE23PyLatheDB,Martins@IEEAccess23Lathe}, a state-of-the-art R-KwS system. Lathe expanded the scope of keyword queries by allowing them to reference both attribute values and schema elements, such as, relation and attribute names. It also introduced an \textit{eager CJN evaluation} mechanism, ensuring that all generated CJNs yield non-empty results, thereby improving the relevance and efficiency of query processing. Despite these advancements, Lathe still struggles to consistently place the most relevant CJNs at the top of the ranking, especially when queries are ambiguous. This limitation highlights the need for more robust and context-aware ranking techniques.

Recent breakthroughs in Natural Language Processing (NLP), particularly transformer-based models, have demonstrated remarkable success in capturing nuanced semantic relationships and contextual cues across various ranking and retrieval tasks. Their adaptability suggests that these models could similarly improve R-KwS systems, where complexity and ambiguity pose persistent challenges. However, integrating transformer-based techniques requires bridging the gap between textual language models and the structured, tabular representations of QMs and CJNs. To address this issue, we propose converting QMs and CJNs into text-like representations through a linearization process, making them suitable inputs for sentence transformers. Moreover, to supply the required training data in a suitable way, we propose a data augmentation strategy for fine-tuning, enabling the model to generalize more effectively and handle diverse, ambiguous queries.

Our approach transforms Lathe’s ranking pipeline, moving beyond heuristic or Bayesian methods. By leveraging transformer-based models, we introduce a richer, more context-sensitive mechanism for identifying the most relevant QMs and CJNs. This approach enhances the overall ranking quality, improves key metrics such as Recall and $R@k$, and ultimately helps overcome Lathe’s prior limitations.

In summary, our main contributions are:
(i) the introduction of a transformer-based ranking approach that enriches Lathe, a state-of-the-art R-KwS system, with a more context-aware understanding of QMs and CJNs;
(ii) the development of a linearization strategy to represent QMs and CJNs as textual sequences suitable for input to sentence transformers;
(iii) the implementation of a data augmentation strategy that improves model fine-tuning and enables better handling of ambiguous queries; and
(iv) empirical evidence showing that our transformer-based ranking significantly outperforms the original Lathe and other advanced R-KwS methods, leading to improved effectiveness in surfacing the most relevant results.

This paper is organized as follows. Section~\ref{chap:related-work} surveys related work on relational keyword search systems and transformer-based approaches for tabular data. Section~\ref{chap:overview} provides an overview of the Lathe framework, including keyword matching, query matching, and CJN generation and ranking steps. In Section~\ref{sec:transformer-qm-ranking}, we present our transformer-based approach for Query Match (QM) ranking, and in Section~\ref{sec:transformer-cjn-ranking}, we detail how transformers are employed to improve Candidate Joining Network (CJN) ranking. Section~\ref{sec:data-agumentation} introduces our data augmentation strategies, designed to enhance the model’s robustness and generalization capabilities. Next, Section~\ref{chap:experiments} reports our experimental results and evaluates the effectiveness of our methods. Finally, Section~\ref{chap:conclusion} concludes by summarizing our contributions and discussing potential avenues for future research.

\section{Background and Related Work}\label{chap:related-work}

In this chapter, we discuss the background and related work on keyword search systems over relational databases, focusing on approaches based on Schema Graphs, and on deep neural networks (DNNs) applied to tabular data. For a more comprehensive view of the state-of-the-art in keyword-based and natural language queries over databases, we refer the interested reader to a survey on this matter~\cite{Affolter@VLDBJ2019_SurveyNLIDB}.

\subsection{Relational Keyword Search Systems}

Current Relational Keyword Search (R-KwS) systems generally fall into two distinct categories: those based on \emph{Schema Graphs} and those based on \emph{Instance Graphs}.

\subsubsection*{Instance Graph-based Systems.} Systems like BANKS \cite{Aditya@VLDB02BANKS}, BANKS-II \cite{Kacholia@VLDB05Bidirectional}, BLINKS \cite{He@SIGMOD07BLINKS}, and Effective \cite{Liu@SIGMOD06Effective} model the database as an \emph{Instance Graph}, where nodes represent tuples and edges connect tuples related by foreign keys. These systems answer keyword queries by finding small subgraphs (or subtrees) that contain all the query keywords, minimizing some distance measure. They typically merge the tuple-retrieval and answer-structure steps, returning results in one go. While this approach can be intuitive, it often ignores the explicit structural information provided by the database schema and can incur higher computational costs due to materialization or graph-based indexing overheads.

\subsubsection*{Schema Graph-based Systems.} Systems in this category rely on the database schema to guide the search process. They generate \emph{Candidate Joining Networks} (CJNs), which are networks of relations connected by join conditions, and translate these CJNs into SQL queries. The execution of these queries returns \emph{Joining Networks of Tuples} (JNTs), which constitute the final results delivered to the user. This approach was initially proposed by DISCOVER \cite{Hristidis@VLDB02DISCOVER} and DBXplorer \cite{Agrawal@ICDE02DBXplorer} and was subsequently adopted by DISCOVER-II \cite{Hristidis@VLDB03Efficient}, SPARK \cite{Luo@SIGMOD07Spark}, CD \cite{Coffman@KEYS10CD}, KwS-F \cite{Baid@VLDB10KwSF}, CNRank \cite{Oliveira@ICDE15CNRank}, MatCNGen \cite{Oliveira@ICDE18MatCNGen,Oliveira@TKDE20}, and Lathe.

The best-known algorithm for CJN generation, CNGen, was introduced in DISCOVER~\cite{Hristidis@VLDB02DISCOVER} and later adopted as default in most R-KwS systems. Although CNGen guarantees a complete, non-redundant set of CJNs, it often produces a large number of them, leading to high computational costs in generation and evaluation. Subsequent works addressed these challenges in various ways. DISCOVER-II \cite{Hristidis@VLDB03Efficient}, SPARK \cite{Luo@SIGMOD07Spark}, and CD \cite{Coffman@KEYS10CD} introduced IR-style scoring functions to rank JNTs, improving the relevance of top-K results. KwS-F~\cite{Baid@VLDB10KwSF} tackled scalability by imposing time limits and allowing partial results. CNRank~\cite{Oliveira@ICDE15CNRank} reduced evaluation costs by ranking CJNs themselves, thus evaluating only the most promising candidates. MatCNGen~\cite{Oliveira@ICDE18MatCNGen,Oliveira@TKDE20} further optimized CJN generation by considering QMs, significantly decreasing the total number of CJNs and improving performance.

Lathe~\cite{Martins@ICDE23PyLatheDB,Martins@IEEAccess23Lathe}, building upon these advances, offered a more efficient CJN generation process and introduced an eager CJN evaluation mechanism, ensuring that all evaluated CJNs return non-empty results. Although Lathe represents the state-of-the-art in this lineage, it still encounters difficulties in consistently identifying and ranking the most relevant CJNs at the top—especially in ambiguous query scenarios. This limitation highlights the need for context-aware ranking methods that capture subtle relationships in the data and query structures.

Coffman \& Weaver~\cite{Coffman@CIKM10Framework} proposed a framework for evaluating and comparing R-KwS systems, providing standardized datasets (MONDIAL, IMDb, Wikipedia) and query workloads. This framework has been widely adopted in subsequent studies on R-KwS~\cite{Luo@SIGMOD07Spark,Oliveira@ICDE15CNRank,Oliveira@ICDE18MatCNGen,Oliveira@TKDE20,Coffman@TKDE12Evaluation}, including those evaluating Lathe and its predecessors.

\subsection{Deep Neural Networks and Tabular Data}

Deep neural networks (DNNs) have revolutionized the field of machine learning, achieving remarkable success in various domains such as images, audio, and text~\cite{Devlin@arXiv18BERT}. However, their application to tabular data, which is characterized by a mix of numerical and categorical features, presents unique challenges. The heterogeneous nature of tabular data complicates the direct application of DNNs, prompting the development of specialized approaches to better handle this data type~\cite{shwartz@InfoFusion22Tabular}.

Recent advancements in DNNs for tabular data can be categorized into three main groups: data transformations, specialized architectures, and regularization models~\cite{borisov@ArXiv22DDNSurvey}. Data transformation techniques convert tabular data into formats more suitable for neural networks, often employing sophisticated encoding strategies for categorical variables. Specialized architectures, including hybrid models and transformer-based approaches, are designed to leverage the unique characteristics of tabular data. Regularization models focus on preventing overfitting and enhancing the generalization capabilities of DNNs.

Among these, specialized architectures form the largest group of approaches. Transformer-based approaches~\cite{vaswani2017attention} have been adapted for tabular data, using attention mechanisms to handle diverse features~\cite{arik2021tabnet}.

An important problem in this domain is table retrieval, which involves identifying the most relevant table from a set of tables given a specific query~\cite{zhang@WWW18AdHocTableRetrieval, zhang@TWEB21TableRetrievalSemantic}. This task is crucial for finding tables that can answer a given question or provide relevant information. Considering that Candidate Joining Networks (CJNs) can be interpreted as database views, the ranking of CJNs can be seen as a variant of the table retrieval problem. Both tasks require understanding and matching the structure and content of tables or views to a given query or information need.

Several transformer-based systems have been developed to handle table retrieval effectively~\cite{badaro@TACL23TransformersTabular}. TaBERT~\cite{Yin@arXiv20TaBERT} integrates natural language text with structured tabular data by linearizing table structures and introducing content snapshots. Linearization is necessary because transformer models are traditionally designed to process unstructured data like text, so the structured rows and columns of a table must be converted into a serialized sequence format. StruBERT~\cite{Trabelsi@arXiv22StruBERT} builds on TaBERT by incorporating horizontal self-attention, leading to improvements in table retrieval and similarity tasks. TAPAS~\cite{Herzig@ArXiv20TAPAS} extends the BERT model to jointly encode table structures and questions, facilitating various operations directly within the table context.

While table retrieval systems like TaBERT, StruBERT, and TAPAS primarily handle single tables, CJNs span different database tables, making the retrieval and ranking tasks more complex. This complexity highlights the need for specialized approaches to manage multi-table scenarios effectively, pushing the boundaries of what current transformer-based systems can achieve.

More recently, models like Table-GPT, TableLlama, and TableLLM have been proposed to advance the state of tabular data processing by leveraging large language models (LLMs). Table-GPT~\cite{li2023table} adapts the GPT architecture specifically for table tasks, demonstrating significant improvements in tasks such as table question answering, table-to-text generation, and table-based data augmentation. TableLlama~\cite{zhang2023tablellama} focuses on creating open, large generalist models that handle a wide variety of table-related tasks. TableLLM~\cite{zhang2024tablellm} targets real office usage scenarios, enabling LLMs to perform complex tabular data manipulations, including table joining, filtering, and aggregation.

Despite these advancements, LLMs like Table-GPT, TableLlama, and TableLLM face limitations. One significant drawback is their computational complexity and resource requirements, often requiring substantial data and computational power for training and inference, which can be impractical for many real-world applications. Additionally, while LLMs excel at handling specific tasks they are fine-tuned for, they may struggle with tasks involving intricate relationships across multiple tables, such as those required for CJN ranking. The integration of multiple data sources and the complexity of understanding the combined semantics of these sources remain challenging for current LLMs, highlighting the need for further advancements in this area.

\section{ {\metodo}  Overview}\label{chap:overview} 

This section provides an overview of the {\metodo} system, which was previously presented  elsewhere~\cite{Martins@ICDE23PyLatheDB,Martins@IEEAccess23Lathe}. Here, we outline the existing challenges in the original version of {\metodo} and introduce the solutions we propose in this paper to address these issues. These enhancements aim to improve both the accuracy and efficiency of the system, advancing its utility for relational keyword search.

To guide our discussion, we illustrate in Figure~\ref{fig:imdb_instance} a simplified excerpt from the well-known IMDB\footnote{Internet Movie Database https://www.imdb.com/interfaces/}.

\begin{figure}[htb]
	\centering
	\setlength\fboxsep{10pt} 
	\begin{adjustbox}{max width=\textwidth}			
		\begin{tabular}{llll}

\begin{tabular}{|c|l|}
    \multicolumn{2}{@{}l}{\textbf{PERSON}} \\
    \hline
    \textbf{ID} & \textbf{Name} \\ 
    \hline
    1 & Will Smith \\ 
    2 & Will Theakston \\ 
    3 & Maggie Smith \\ 
    4 & Sean Bean \\ 
    5 & Elijah Wood \\ 
    6 & Angelina Jolie \\ 
    \hline
\end{tabular}

			&
			\multicolumn{3}{l}{
\begin{tabularx}{0.8\textwidth}{|c|X|c|}
    \multicolumn{3}{@{}l}{\textbf{MOVIE}} \\
    \hline
    \textbf{ID} & \textbf{Title} & \textbf{Year} \\ 
    \hline
    7 & Men in Black & 1997 \\
    8 & I am Legend & 2007 \\
    9 & Harry Potter and the Sorcerer's Stone & 2001 \\
    10 & The Lord of the Rings: The Fellowship of the Ring & 2001 \\
    11 & The Lord of the Rings: The Return of the King & 2003 \\
    12 & Mr. \& Mrs. Smith & 2005 \\
    \hline
\end{tabularx}}
			\\
			\\
            \multicolumn{2}{l}{
\begin{tabular}{|c|l|}
    \multicolumn{2}{@{}l}{\textbf{CHARACTER}} \\
    \hline
    \textbf{ID} & \textbf{Name} \\ 
    \hline
    13 & Agent J \\
    14 & Robert Neville \\
    15 & Marcus Flint \\
    16 & Minerva McGonagall \\
    17 & Boromir \\
    18 & Frodo Baggins \\
    19 & Jane Smith \\
    \hline
\end{tabular}}
			&
			\begin{tabular}{|c|l|}
    \multicolumn{2}{@{}l}{\textbf{ROLE}} \\
    \hline
    \textbf{ID} & \textbf{Name} \\ 
    \hline
    20 & Actor \\
    21 & Actress \\
    22 & Producer \\
    23 & Writer \\
    24 & Director \\
    25 & Editor \\
    \hline
\end{tabular}
			& 
			\begin{tabular}{|c|c|c|c|c|}
    \multicolumn{5}{@{}l}{\textbf{CASTING}} \\
    \hline
    \textbf{ID} & \textbf{PID} & \textbf{MID} & \textbf{ChID} & \textbf{RID} \\
    \hline
    26 & 1 & 7  & 13 & 20 \\  
    27 & 1 & 8  & 14 & 20 \\  
    28 & 2 & 9  & 15 & 20 \\  
    29 & 3 & 9  & 16 & 21 \\  
    30 & 4 & 10 & 17 & 20 \\  
    31 & 4 & 11 & 17 & 20 \\  
    32 & 5 & 10 & 18 & 20 \\  
    33 & 5 & 11 & 18 & 20 \\  
    34 & 6 & 12 & 19 & 21 \\  
    \hline
\end{tabular}
		\end{tabular}
	\end{adjustbox}	
	\caption{A simplified excerpt from IMDB}
	\label{fig:imdb_instance}
\end{figure}


Consider a user keyword query $Q{=}{``will\,smith\,films"}$  to list the movies in which Will Smith appears. Intuitively, the terms ``\textit{will}'' and ``\textit{smith}'' are likely to match the contents of a relation from the DB, whereas ``films'' may match the name of a relation or attribute. 

As other previous methods in the literature (e.g., CNGen~\cite{Hristidis@VLDB02DISCOVER} and MatCNGen \cite{Oliveira@ICDE18MatCNGen,Oliveira@TKDE20}), the main goal of  {\metodo} is, given a query such as $Q$, generating a SQL query that,
when executed, fulfills the information needed for the user. The difference between {\metodo} and these previous methods is that they are not able to handle references to schema elements, such as ``films'' in $Q$.

For query $Q$, two of the possible SQL queries that would be generated are presented in Figures~\ref{fig:imdb_sample_sql_queries}~(a) ($S_1$) and~(b) ($S_2$), whose respective results for the database of Figure~\ref{fig:imdb_instance} are presented in Figures~\ref{fig:imdb_sample_sql_queries}(c) and~(d). In the query $S_1$, the keywords "will" and "smith" match the value of a single tuple of relation PERSON, while the keyword "films" matches the name of the relation MOVIE. As a result, $S_1$ retrieves the movies which the person Will Smith was in, and thus, satisfies the original user intent. As for query $S_2$, the keywords "will" and "smith" match values of two different tuples in relation PERSON, that is, they refer to two different persons. The keyword "films" matches the name of the relation MOVIE again. Therefore, $S_2$ retrieves movies in which two different persons, whose names respectively include the terms ``\textit{will}'' and ``\textit{smith}'', participated in. In these case, the persons are Will Theakston and Maggie Smith.

\begin{figure}[htb]
	\newsavebox{\mylistingbox}
	\centering
    \vspace{-5mm}
	\begin{lrbox}{\mylistingbox}
		\begin{tabular}{cc}
	\begin{lstlisting}
	SELECT m.title, p.name
	FROM person p
	JOIN casting c ON  p.id=c.person_id
	JOIN movie m ON m.id = c.movie_id
	WHERE p.name ILIKE '%will%' 
	AND p.name ILIKE '%smith%';
	\end{lstlisting} 
	&
	\begin{lstlisting}
	SELECT m.title, p1.name, p2.name
	FROM person p1
	JOIN casting c1 ON p1.id=c1.person_id
	JOIN movie m ON m.id = c1.movie_id
	JOIN casting c2 ON m.id = c2.movie_id
	JOIN person p2 ON p2.id=c2.person_id
	WHERE p1.name ILIKE '%will%' 
	AND p2.name ILIKE '%smith%'
	AND p1.id<>p2.id;
	\end{lstlisting}
	\\
	(a) & (b) \\\\
	\begin{tabular}{|l|l|} 
		\hline
		\textbf{m.title} & \textbf{p.name}  \\ 
		\hline
		Men in Black     & Will Smith       \\
		\hline
		I am Legend      & Will Smith       \\
		\hline
	\end{tabular}
	&
	\begin{tabular}{|p{3.5cm}|l|l|} 
		\hline
		\textbf{m.title}                      & \textbf{p1.name} & \textbf{p2.name}  \\ 
		\hline
		Harry Potter and the Sorcerer's Stone & Will Theaskton   & Maggie Smith      \\
		\hline
	\end{tabular}
	\\
	(c) & (d) \\
\end{tabular}
	\end{lrbox}
	\scalebox{0.7}{\usebox{\mylistingbox}}
	\caption{SQL queries generated for the keyword query ``\textit{will smith movies}'' and their returned results.}
	\label{fig:imdb_sample_sql_queries}
\end{figure}

	

As this example indicates, there may be several plausible SQL queries related to a given keyword query. Therefore, it is necessary to decide which alternative is more likely to fulfill the user intent. This task is also carried out by \metodo. 

Next, we present an overview of the components and the functioning of \metodo.

\subsection{System Architecture} 
\label{sec:architecture}

\begin{figure*}[!htb]
	\centering
	\includegraphics[width=\textwidth]{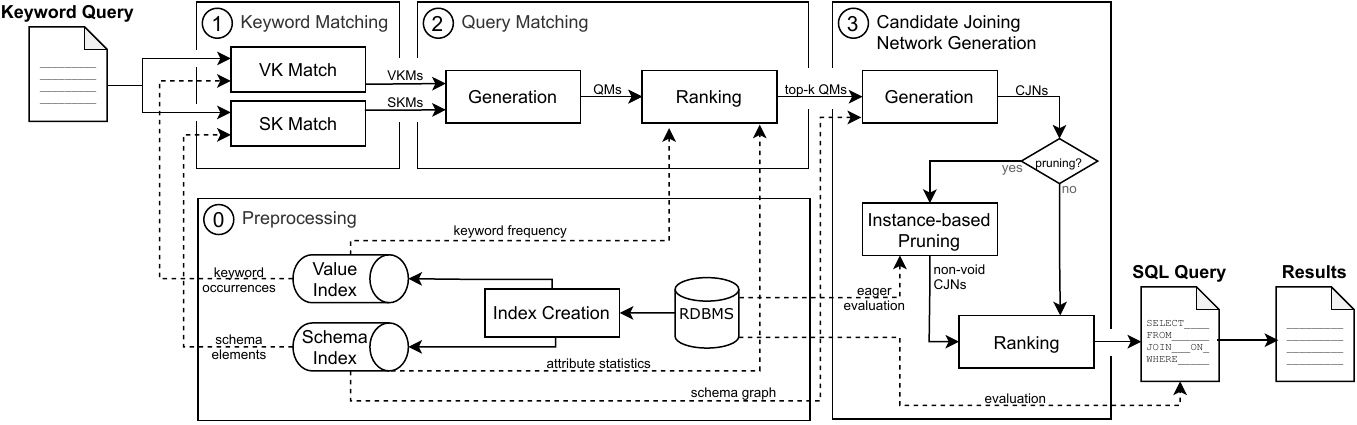}
	\caption{Main phases and architecture of {\metodo}}
	\label{fig:lathe_arch}
\end{figure*}

Figure~\ref{fig:lathe_arch} illustrates the main phases that comprise the operation of \metodo.
The process begins with an input keyword query posed by the user and includes the steps described in the following. 

\paragraph*{Preprocessing~\circled{0}}
Prior to the execution of the keyword queries, {\metodo} performs a setup step, in which the system creates two data structures: the \textit{Value Index} and the \textit{Schema Index}, which respectively, store information about the values of the attributes and the database schema elements. These indexes are used to retrieve keyword matches, score QMs and CJNs, and build the graph of the database schema. Typically, the indexes do not need to be updated often.

\paragraph*{Keyword Matching~\circled{1}}
In this step, the system first performs a data cleansing on the query, removing stopwords, punctuation, and changing capitalization.
Then, it attempts to associate each of the keywords from the query with a database schema element based on the names of relations and attributes, or the values of the attributes.
The system then generates sets of  \textit{Value-Keyword Matches} (VKMs), which associate keywords with sets of tuples whose attribute values contain these keywords, and \textit{Schema-Keyword Matches} (SKMs), which associate keywords with names of relations or attributes based on similarity functions from the lexical database WordNet~\cite{Miller@98Wordnet}.
Using the previously constructed \textit{Value Index} and \textit{Schema Index}, these matches are efficiently retrieved, allowing the system to map keywords to relevant tuples and schema elements quickly without further interaction with the DBMS. 
We represent a keyword match using the following notation: $R^X[A_1^{K_1},\ldots,A_n^{K_n}]$, where $R$ is a database relation, $A_i$ is an attribute from $R$, $K_i$ is the set of keywords associated with $A_i$, and $X \in \{X,V\}$ indicates if the association targets schema elements or attributes values, respectively.

In Table \ref{tab:keyword_match} we show possible matches between keywords in the input query and the database elements.
For example, the keywords ``will smith'' are found together in the values of the attribute \texttt{name} of the \texttt{PERSON} relation. 
The keyword ``will'' is also found alone in the values of \texttt{PERSON.name}, which is the case of the person Will Theakston present in instance shown in Figure~\ref{fig:imdb_instance}.
The term ``\emph{smith}'' can refer to either the name of a person, the name of a character or even the title of a movie, in this case ``Mr. \& Mrs. Smith''.
Since these keywords are part of attribute values, these matches are considered VKMs. In the case of the keyword ``\emph{films}'', it actually matches the \emph{name} of the \texttt{Movie} relation, which is why in Table~\ref{tab:keyword_match} the keyword ``\emph{films}'' matches \texttt{MOVIE.self}. Thus, this match is considered an SKM.

\begin{table}[htb!]
    \centering
    \caption{Keyword matched for the query "\textit{will smith films}"}
        \renewcommand{\arraystretch}{1.5}
        \footnotesize
        \begin{tabular}{|c|c|c|l|l|}
            \hline
            \textbf{Keywords}              
            & \textbf{Type} 
            & \textbf{Database Element} 
            & \textbf{Relational Algebra}
            & \textbf{Keyword Match}
            
            \\ \hline
            \textit{will smith}            
            & value         
            & \texttt{PERSON.name}  
            & $\sigma_{\texttt{name} \supseteq \{will,smith\} }(\texttt{PERSON})$ 
            & $PERSON^V[name^{ \{will,smith\} }]$
            \\ \hline
            
            \textit{will}                 
            & value         
            & \texttt{PERSON.name}   
            & $\sigma_{\texttt{name} \ni will}(\texttt{PERSON})$    
            & $PERSON^V[name^{ \{will\} }]$        
            
            \\ \hline
            \multirow{3}{*}{\textit{smith}} 
            & value         
            & \texttt{PERSON.name}  
            & $\sigma_{\texttt{name} \ni smith}(\texttt{PERSON})$  
            & $PERSON^V[name^{ \{smith\} }]$            
            \\ \hhline{~|----} 
                                                    
            & value         
            & \texttt{CHARACTER.name}  
            & $\sigma_{\texttt{name} \ni smith}(\texttt{CHARACTER})$    
            & $CHARACTER^V[name^{ \{smith\} }]$      
            \\ \hhline{~|----}  
                                                    
            & value         
            & \texttt{MOVIE.title}   
            & $\sigma_{\texttt{title} \ni smith}(\texttt{MOVIE})$
            & $MOVIE^V[title^{ \{smith\} }]$
            \\ \hline
                        
            \textit{films}                 
            & schema        
            & \texttt{MOVIE.self}        
            & \texttt{MOVIE}
            & $MOVIE^S[self^{ \{films\} }]$
            \\ \hline
        \end{tabular}
        \renewcommand{\arraystretch}{1}
    \label{tab:keyword_match}
\end{table}

\paragraph*{Query Matching~\circled{2}}
Second, {\metodo} combines keyword matches (VKMs and SKMs) so that they form a \textit{total} and \textit{minimal} cover for the keyword query.
That is, the keyword matches comprise all the keywords from the query and no keyword match is redundant. 
These keyword match combinations are referred to as \textit{Query Matches} (QMs). 

All possible QMs of the KMs illustrated in Table~\ref{tab:keyword_match} are listed below:
\begin{alignat*}{1}
	&M_1 = \{PERSON^V[name^{ \{will,smith\} }], MOVIE^S[self^{ \{films\} }]\}\\
	&M_2 = \{PERSON^V[name^{ \{will\} }],PERSON^V[name^{ \{smith\} }] , MOVIE^S[self^{ \{films\} }]\}\\
	&M_3 = \{
	PERSON^V[name^{ \{will\} }],
    CHARACTER^V[name^{ \{smith\} }],
	MOVIE^S[self^{ \{films\} }]
	   \}\\
	&M_4 = \{PERSON^V[name^{ \{will\} }], MOVIE^V[title^{ \{smith\} }], MOVIE^S[self^{ \{films\} }]\}
\end{alignat*}

Although, in general, there may be a large number of QMs due to their combinatorial nature, only a few are useful in producing plausible answers that align with the user’s intention. In our example, the query match $M_1$ above would best represent the intention of the user.

To efficiently handle this, {\metodo} employs a ranking algorithm to prioritize only the top-ranked QMs for further processing. In previous work, this algorithm relied on classic features commonly used in Information Retrieval, combined using a Bayesian Network~\cite{Mesquita@IPM07LABRADOR,Oliveira@ICDE18MatCNGen,Oliveira@TKDE20,Martins@IEEAccess23Lathe,Martins@ICDE23PyLatheDB}, which proved effective in selecting relevant QMs within the top-ranking positions~\cite{Oliveira@TKDE20,Martins@IEEAccess23Lathe}. However, the transformer-based approach introduced in this paper further improves the ranking quality and alignment with user intent, leveraging the transformer models capability in exploring the implicit semantics in relational data~\cite{Yin@arXiv20TaBERT,Trabelsi@arXiv22StruBERT}. Details of this approach are provided in Section~\ref{sec:transformer-qm-ranking}.


\paragraph*{Candidate Joining Network Generation~\circled{3}}

Lastly, the system generates possible interpretations for the keyword query. That is, the system tries to connect all the keyword matches from the QMs through CJNs, which are based on the graph of referential integrity constraints (PK/FK) of the target database schema.

CJNs can be thought of as relational algebra joining expressions that translate directly into SQL queries. For instance, both QMs $M_1$ and $M_2$ defined above can be connected using the \texttt{CASTING} relation, resulting in CJNs whose SQL translations are presented in Figures~\ref{fig:imdb_sample_sql_queries}\,(a) and~(b), respectively.

Currently, {\metodo} relies on the ranking of QMs as a foundational step in determining the most relevant CJNs, and consequently on the Bayesian network model. Previous studies~\cite{Oliveira@TKDE20,Martins@IEEAccess23Lathe} have demonstrated that this algorithm is highly effective in selecting the most relevant CJN within the top-5 ranking positions. However, it often struggles to identify the best CJN in the top-1 position.

To address this limitation, in this paper, we introduce a transformer-based ranking approach for CJNs that leverages transformer models to assess and rank CJNs based on their semantic relevance to the keyword query. This approach enables a deeper understanding of semantic relevance by allowing the model to prioritize CJNs in a way that aligns more closely with the user’s intended meaning. Recent advancements in applying transformer models to relational keyword search tasks have demonstrated the potential for improved relevance and accuracy in such applications~\cite{Devlin@arXiv18BERT, Reimers@arXiv19SBERT, Yin@arXiv20TaBERT, Trabelsi@arXiv22StruBERT}. Building on these advancements, our transformer-based approach enhances CJN ranking by embedding relational data in a format that can be effectively processed by neural models, resulting in more accurate and relevant search outcomes. The details of our proposed ranking approach are presented in Section~\ref{sec:neural-cjnranking}.



Once we have identified the most likely CJNs, they can be evaluated as SQL queries executed by a relational database management system (RDBMS) to present the results to the user. 
Additionally, to prune spurious CJNs (those yielding empty results), {\metodo} performs a preliminary evaluation phase, using an instance-based pruning strategy that assesses these CJNs early and excludes those that lack query relevance.

In the following sections, we present in detail our main technical contributions in this paper. 
Section~\ref{sec:transformer-qm-ranking} presents the Transformer-based Query Match (QM) ranking, while Section~\ref{sec:transformer-cjn-ranking} describes the Transformer-based Candidate Joining Network (CJN) ranking. We then outline our empirical assessment in Section~\ref{chap:experiments}, followed by the Conclusion and insights for future work in Section~\ref{chap:conclusion}.

\section{Transformer-based Query Match Ranking}\label{sec:transformer-qm-ranking}

Our newly proposed algorithm ranks the Query Matches (QMs) for a given keyword query using transformer-based models that assess relevance through semantic similarity. This approach ensures that the most relevant QMs are prioritized, enhancing the accuracy and relevance of search results.

\subsection{QM Linearization}\label{sec:qm-linearization}

To enable compatibility with transformer-based models, we propose a strategy to linearize Query Matches (QMs) from their original tabular format into structured sentences. This transformation allows the QMs, which contain relational data, to be encoded effectively by sentence-transformer models, facilitating semantic similarity comparisons essential for ranking. This is in line with previous work on representing relational data within neural language models\cite{Yin@arXiv20TaBERT,Trabelsi@arXiv22StruBERT}.

The linearization process begins with the keyword query itself, represented as a sentence in the format ``query: $Q$'', where $Q$ is the keyword query. This sentence is then encoded using a sentence-transformer model to generate an embedding for the query. Figure~\ref{fig:encoding-qms} provides an illustration of this process using the example query ``Will Smith films'' and the corresponding QMs.

In this figure, $M_1$ and $M_2$ represent query matches for the example query, which are translated into standardized structured sentences. Each keyword mapping follows the format "table.attribute.type: keywords", with individual components separated by a pipe symbol to ensure consistency with transformer models. This structure enhances the compatibility of QMs with the model, as shown in Example~\ref{ex:qm-encoding}.

\begin{example}
	\label{ex:qm-encoding}
	Considering the query matches $M_1$ and $M_2$ generated previously for the query ``Will Smith films'', their sentences are represented as follows in Figure~\ref{fig:encoding-qms}:

    \begin{figure}[!htb]
        \centering
        \begin{adjustbox}{max width=0.9\textwidth}	
            {\renewcommand{\arraystretch}{1.5} 
            \begin{tabular}{|c|l|}
            \hline
            $Q$         & Will Smith films       \\ 
            \hline
            $sentence(Q)$         & query: Will Smith films       \\ 
           \hline
            $M_1$         & $\{PERSON^V[name^{ \{will,smith\} }], MOVIE^S[self^{ \{films\} }]\}$       \\ \hline
            $sentence(M_1)$ & answer: person.name.value: will smith | movie.self.schema: films                      \\ \hline
            $M_2$         & $\{PERSON^V[name^{ \{will\} }],PERSON^V[name^{ \{smith\} }] , MOVIE^S[self^{ \{films\} }]\}$       \\ \hline
            $sentence(M_2)$ & answer: person.name.value: will | person.name.value: smith | movie.self.schema: films \\ \hline
            \end{tabular}
            }
        \end{adjustbox}
        \caption{Sentence translation of query matches.}
        \label{fig:encoding-qms}
    \end{figure}
\end{example}

This linearization of QMs ensures that the structured relational data is represented in a format that can be processed by transformer models, enabling accurate and efficient similarity-based ranking.

\subsection{QM Ranking Algorithm}

The QM ranking process utilizes sentence-transformer models to rank QMs for a given keyword query. This systematic process is detailed in Algorithm~\ref{alg:neural-qmrank}, and it involves linearization, embedding generation, and similarity-based ranking.

    
    
	

\begin{algorithm}[!htb]
	\caption{NeuralQMRank($Q$,$Q\!M$)}
	\label{alg:neural-qmrank}
	\KwIn{A keyword query $Q$ \newline
        A set of query matches $Q\!M$\;
        }
	\KwOut{The set of ranked query matches $R\!Q\!M$}
    $R\!Q\!M \leftarrow [\ ]$\;        
    \Let{$Model$ be the sentence-transformer model} 
    $S_Q \leftarrow sentence(Q)$\; \label{line:neuralqm-query-sentence-gen}
    $E_Q \leftarrow Model.encode( S_Q )$\;  \label{line:neuralqm-query-encode}
	\For{$M_i \in Q\!M$}{
        $S_{M_i} \leftarrow sentence(M_i)$\; \label{line:neuralqm-qm-sentence-gen}
        $E_{M_i} \leftarrow Model.encode(S_{M_i})$\; \label{line:neuralqm-qm-encode}

        $score \leftarrow sim(E_Q,E_{M_i})$\; \label{line:neuralqm-begin-calculate-score}
        $R\!Q\!M$.append($\langle score, M_i \rangle$)\; 
        \label{line:neuralqm-end-calculate-score}
    }
    \textbf{Sort }$R\!Q\!M$ in descending order\; 
    \label{line:neuralqm-sorting}
	\Return{$R\!Q\!M$}	
\end{algorithm}

First, the algorithm linearizes both the keyword query and the QMs into structured sentences (Lines~\ref{line:neuralqm-query-sentence-gen} and \ref{line:neuralqm-qm-sentence-gen}), as described in Section~\ref{sec:qm-linearization}. This step ensures that the information contained within the keyword query and the QMs is properly structured and represented in a textual format compatible with subsequent analysis.

Next, the algorithm utilizes a sentence-transformer model to generate embeddings for both the linearized keyword query and the linearized QMs (Lines~\ref{line:neuralqm-query-encode} and \ref{line:neuralqm-qm-encode}). These embeddings encapsulate the semantic representations of the sentences, facilitating meaningful comparison and analysis.
The subsequent step involves computing the similarity between the embedding representing the keyword query and the embeddings representing the QMs to determine the relevance of each QM to the keyword query (Lines ~\ref{line:neuralqm-begin-calculate-score}-\ref{line:neuralqm-end-calculate-score}). The similarity measure can be either cosine similarity or dot product similarity, depending on the specific sentence-transformer model used.
Finally, the algorithm sorts the QMs in descending order based on their similarity  (Line~\ref{line:neuralqm-sorting}). This ranking ensures that the most semantically aligned QMs with the keyword query are prioritized, thus improving the relevance of the search results returned to the user.

Initially trained on extensive general corpora, sentence-transformer models are effective at discerning semantic similarities. However, due to their generalized nature, these models benefit from further adaptation to specific tasks, such as QM ranking. Research has shown that fine-tuning on domain-specific data can significantly improve model performance for targeted applications~\cite{Reimers@arXiv19SBERT,Devlin@arXiv18BERT}. Therefore, in the next section, we describe our approach to fine-tuning these models for QM ranking, allowing them to better capture the relevant patterns and relationships in the linearized QM data and enhancing their ranking accuracy.

\subsection{QM Ranking Fine-tuning} \label{sec:qm-finetuning}

In this section, we present our approach to fine-tuning sentence-transformer models for effective QM ranking. In our fine-tuning process, we create training examples composed of tuples that include the keyword query representation, the QM representation, and a similarity score between them. We determine the relevance score of each QM using a Bayesian model, as detailed in previous work \cite{Martins@ICDE23PyLatheDB,Martins@IEEAccess23Lathe,Oliveira@ICDE15CNRank}.
QMs deemed relevant to the keyword query are assigned a score of 1, while other QMs receive a score calculated by applying a sigmoid function to the Bayesian score, with an additional weight of 0.4 to account for negative examples. The similarity score is computed as follows:

\[
    Sim(Q,M)= 
\begin{dcases}
    1, & \text{if } M \text{ is relevant for } Q\\
    \dfrac{1}{1 + e^{-0.4{\times}bayesian\_score(Q,M)}}, & \text{otherwise}
\end{dcases}
\]

where $bayesian\_score(Q,M)$ represents the Bayesian model score for the query match $M$ given the query $Q$.

\begin{example}
	\label{ex:qm-finetuning}
	Considering the sentences for the query matches $M_1$ and $M_2$, as shown in Figure~\ref{fig:encoding-qms}, we generate the following training examples, displayed in Figure~\ref{fig:qms-train-examples}:

    \begin{figure}[!htb]
        \centering
        \begin{adjustbox}{max width=0.7\textwidth}	
            {\renewcommand{\arraystretch}{1.5} 
            \begin{tabular}{|c|p{10cm}|}
            \hline
            Positive Example         & ("query: Will Smith films", "answer: person.name.value: will smith | movie.self.schema: films", score=1.0) \\ \hline
            Negative Example          & ("query: Will Smith films", "answer: person.name.value: will | person.name.value: smith | movie.self.schema: films", score=0.22) \\ \hline
            \end{tabular}
            }
        \end{adjustbox}
        \caption{Training examples for the QM ranking fine-tuning}
        \label{fig:qms-train-examples}
    \end{figure}
\end{example}

Through this fine-tuning process, the model learns to assign higher relevance scores to QMs that closely match the keyword query, thereby improving the quality and relevance of ranked results.

\section{Transformer-based Candidate Joining Network (CJN) Ranking}\label{sec:transformer-cjn-ranking}

While traditional approaches to CJN processing have focused on generation and pruning, we now introduce a neural-based ranking system for CJNs that leverages transformer-based models to assess and rank CJNs based on their semantic relevance to a keyword query. This approach builds upon recent advancements in applying neural language models to relational keyword search tasks~\cite{Devlin@arXiv18BERT, Reimers@arXiv19SBERT,Yin@arXiv20TaBERT,Trabelsi@arXiv22StruBERT}.

Our system ranks Candidate Joining Networks (CJNs) for a given keyword query using transformer-based models, employing a process similar to QM ranking. However, due to the tabular and often multi-row nature of CJNs, the linearization and aggregation steps differ to effectively capture the complex relational information within CJNs.

Since CJNs are generated from relational database tables, a key challenge is to linearize these structured tabular data into text-compatible representations that can be processed by neural language models. This linearization step, detailed in the next section, is crucial for effective CJN ranking.

\subsection{CJN Linearization}\label{sec:cjn-linearization}

Since the results of CJNs, when executed against a database, can be interpreted as database views, we explore techniques designed to encode tabular data. Specifically, we translate each row from the resulting view into a structured sentence and then employ methods to aggregate these sentences into a single representation for the CJN. These aggregation methods capture essential relational nuances embedded within CJNs by consolidating the row-level information retrieved from the database.

To efficiently handle large database views, we adopt a snapshot approach similar to that used in TaBERT~\cite{Yin@arXiv20TaBERT} and StruBERT~\cite{Trabelsi@arXiv22StruBERT}. This technique selects a subset of rows from the database view, ensuring that the data remain within the token capacity constraints of transformer-based models.

Once the snapshot is obtained, we apply one of two primary aggregation techniques to generate a single embedding of the CJN: mean and combination.

\subsubsection*{Mean Approach}

\begin{figure}[!htb]
    \centering
     \includegraphics[width=.95\textwidth]{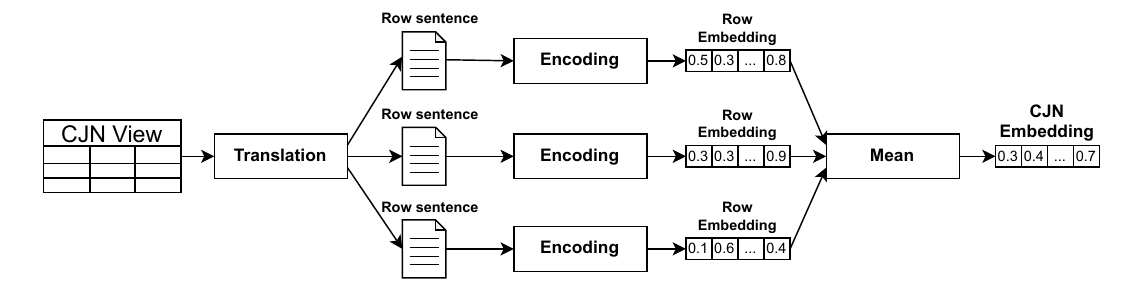}
    \caption{Mean Approach for CJN Linearization}
    \label{fig:mean-cjn-linearization}
\end{figure}

In the mean approach, we encode the sentence for each row individually and then compute the average embedding of all the row embeddings, resulting in a single embedding that represents the entire CJN
Figure~\ref{fig:mean-cjn-linearization} presents the steps for the mean approach.
The sentences generated for CJNs using this approach are shown in Example~\ref{ex:cjn-encoding}.

\subsubsection*{Combination Approach}

\begin{figure}[!htb]
    \centering
     \includegraphics[width=.95\textwidth]{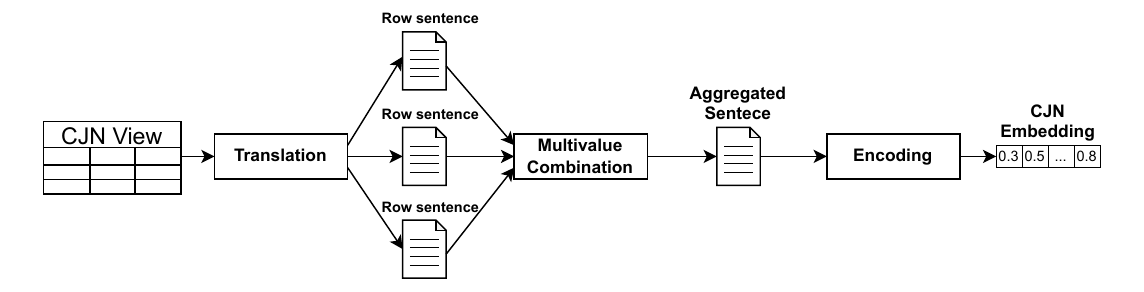}
    \caption{Combination Approach for CJN Linearization}
    \label{fig:combination-cjn-linearization}
\end{figure}

The combination approach involves aggregating the information from multiple rows in the snapshot into a single comprehensive sentence, then encoding this aggregated sentence to represent the CJN. Figure~\ref{fig:combination-cjn-linearization} presents the steps for the combination approach. 
This approach aggregates the rows using the Multivalue technique, which concatenates the values of each attribute across all rows within the snapshot, then we generate a single comprehensive sentence.


The sentences generated for CJNs using this approach are also shown in the Example~\ref{ex:cjn-encoding}.

\begin{example}
	\label{ex:cjn-encoding}
	Considering the query matches $M_1$ and $M_2$ encoded in Example~\ref{ex:qm-encoding}, we can generate the candidate joining networks $C\!J\!N_1$ and $C\!J\!N_2$, shown with their respective results in Figure~\ref{fig:cjn-results}. Based on a snapshot, with the three first rows of the results, we are able to generate the sentences shown in Figure~\ref{fig:encoding-cjns}.
 
    \begin{figure}[!htb]
        \centering
        \begin{adjustbox}{max width=\textwidth}	
            \begin{tabular}{cc}
                $\tikz[baseline]{
                \node[main node,anchor=base] (1) {$C\!J\!N_1=PERSON^V[name^{ \{will,smith\} }]$};
                \node[main node] (2) [below = 3mm of 1] {$CASTING$};
                \node[main node] (3) [below = 3mm of 2] {$MOVIE^S[self^{\{films\}}]$};
                \path[draw,thick,->]
                (2) edge node {} (1)
                (2) edge node {} (3)
                }$
                &
                $\tikz[baseline]{
                \node[main node,anchor=base] (1) {$C\!J\!N_2=MOVIE^S[self^{\{films\}}]$};
                \node[main node] (2) [right = 3mm of 1] {$CASTING$};
                \node[main node] (3) [right = 3mm of 2] {$PERSON^V[name^{ \{will\} }]$};
                \node[main node] (4) [below = 3mm of 1] {$CASTING$};
                \node[main node] (5) [right = 3mm of 4] {$PERSON^V[name^{ \{smith\} }]$};
                \path[draw,thick,->]
                (2) edge node {} (1)
                (2) edge node {} (3)
                (4) edge node {} (1)
                (4) edge node {} (5)
                }$
                \\
                \\
                {\renewcommand{\arraystretch}{1.2}
                \begin{tabular}{|l|l|}
                \hline
                \multicolumn{2}{|l|}{
                    \textbf{Results for} $C\!J\!N_1$
                }
                \\ \hline 
                \textbf{t3.title} & \textbf{t1.name} \\\hline
                Bad Boys           & Smith, Will \\ \hline
                Enemy of the State & Smith, Will \\ \hline
                Free Enterprise    & Smith, Will \\ \hline
                Ali                & Smith, Will \\ \hline
                A Closer Walk      & Smith, Will \\ \hline
                \end{tabular}
                }
            
                 &
            
                {\renewcommand{\arraystretch}{1.2}
                \begin{tabular}{|l|l|l|l|}
                \hline
                \multicolumn{4}{|l|}{
                    \textbf{Results for} $C\!J\!N_2$
                }
                \\ \hline
                \textbf{t3.year} & \textbf{t3.title}     & \textbf{t1.name} & \textbf{t5.name}      \\ \hline
                1944             & The Last Horseman     & Wills, Luke      & Smith, Tom            \\ \hline
                1977             & Looking Up            & Hussing, Will    & Smith, Andrew         \\ \hline
                1977             & Who Has Seen the Wind & Woods, Will      & Smith, Cedric         \\ \hline
                1981             & Urgh! A Music War     & Sergeant, Will   & Smith, Barry          \\ \hline
                1999             & The Lost Son          & Welch, Will      & Smith, Rachel Quigley \\ \hline
                \end{tabular}
                }
                
            \end{tabular}
        \end{adjustbox}
        \caption{CJNs and their results returned from the database.}
        \label{fig:cjn-results}
    \end{figure}

    \begin{figure}[!htb]
        \centering
        \begin{adjustbox}{max width=\textwidth}	
            {\renewcommand{\arraystretch}{1.2}
            \begin{tabular}{|>{\centering\arraybackslash}p{3cm}|>{\raggedright}p{9cm}|>{\raggedright\arraybackslash}p{9cm}|}
            \hline
            \textbf{Aggr. Approach}          & \textbf{Sentences for $C\!J\!N_1$}      & \textbf{Sentences for $C\!J\!N_2$}                                            \\ \hline
            \multirow{3}{*}{\textbf{Mean}} & answer: person.name: Smith, Will | movie: films & answer: person.name: Wills, Luke | movie: films | person.name: Smith, Tom      \\ \hhline{~--}
                                            & answer: person.name: Smith, Will | movie: films & answer: person.name: Hussing, Will | movie: films | person.name: Smith, Andrew \\ \hhline{~--}
                                            & answer: person.name: Smith, Will | movie: films & answer: person.name: Woods, Will | movie: films | person.name: Smith, Cedric   \\ \hline
            \textbf{Combination} &
              answer: person.name: Smith, Will, Smith, Will, Smith, Will | movie: films &
              answer: person.name: Wills, Luke, Hussing, Will, Woods, Will | movie: films | person.name: Smith, Tom, Smith, Andrew, Smith, Cedric \\ \hline
            \end{tabular}
            }
        \end{adjustbox}
        \caption{Sentence representations for the candidate joining networks  $C\!J\!N_1$ and $C\!J\!N_2$.}
        \label{fig:encoding-cjns}
    \end{figure}

    It is noteworthy that while the results for $C\!J\!N_1$ exhibit distinct rows, representing them as sentences reveals some duplicates. This occurrence arises because the keyword ``films'' refers to the table name rather than its values. This observation underscores the nuanced challenges in accurately representing CJN results through sentence encoding.

\end{example}

\subsection{CJN Ranking Algorithm}

The neural CJN ranking is carried out by the algorithm detailed in Algorithm~\ref{alg:neural-cjnrank}. To rank the CJNs for a given keyword query, the algorithm first executes queries generated from the CJNs against the database to obtain the database views they return. It then linearizes the rows of these views and generates embeddings, applying aggregation techniques to create a single representation for the CJN.

    
	
%

\begin{algorithm}[!htb]
	\caption{NeuralCJNRank($Q$,$C\!J\!N$)}
	\label{alg:neural-cjnrank}
	\KwIn{A keyword query $Q$\newline
          A set of candidate joining networks $C\!J\!N$\newline
          A literal variable indicating the aggregation approach $agg$\; 
	}
	\KwOut{The set of ranked candidate joining networks $R\!C\!J\!N$}
    
    $R\!C\!J\!N \leftarrow [\ ]$\;    
    \Let{$Model$ be the sentence-transformer model}
    $S_Q \leftarrow sentence(Q)$\; 
    $E_Q \leftarrow Model.encode(S_Q)$\; \label{line:neuralcjn-query-embedding}
	\For{$C \in C\!J\!N$}{
        \Let{$view$ be the resulting view when running $C$ against the database.} \label{line:neuralcjn-get-views}
        \If{$agg =$ "mean"}{ \label{line:neuralcjn-begin-linearization}
            $E_{view} \leftarrow \{\ \}$\;
            \For{$row \in view$}{
                $S_{row} \leftarrow row\_sentence(row)$\;
                $E_{row} \leftarrow Model.encode(S)$\;
                $E_{view} \leftarrow E_{view} \cup E_{row}$\;                
            }
            $E_{C} \leftarrow mean(E_{view})$\;
        }
        \Else{
            $S_{view} \leftarrow multivalue\_sentence(view)$\;
            $E_{C} \leftarrow Model.encode(S_{view})$\;
        } \label{line:neuralcjn-end-linearization}
        $score \leftarrow sim(E_Q,E_{C})$\; \label{line:neuralcjn-begin-scoring}
        $R\!C\!J\!N$.append($\langle score, C \rangle$)\; \label{line:neuralcjn-end-scoring}
    }	
	\textbf{Sort }$R\!C\!J\!N$ in descending order\; \label{line:neuralcjn-sorting}
	
	\Return{$R\!C\!J\!N$}	
\end{algorithm}

Finally, the algorithm computes the similarity between the keyword query's embedding and the CJN's aggregated embedding to rank the CJNs based on relevance. The ranked CJNs are then returned as the final output.

\subsection{CJN Ranking Fine-tuning}\label{sec:cjn-finetuning}

To effectively rank CJNs, we fine-tune sentence-transformer models to better capture the unique patterns in relational data. Due to the generalized nature of these pre-trained models, they often lack domain-specific weights needed for CJN ranking in relational data contexts. However, state-of-the-art research has demonstrated that fine-tuning significantly enhances the performance of language models on specialized tasks and domains \cite{Reimers@arXiv19SBERT,Devlin@arXiv18BERT}. By fine-tuning pre-trained sentence-transformer models with our training set, we enable them to capture the unique relational nuances required for CJN ranking, thus aligning the model more closely with the complexities of our task.

In the fine-tuning process for CJN ranking, sentence-transformer models are adapted using specific training examples. As CJNs can generate several sentences, multiple training examples may be created for a single CJN. Each example consists of a tuple comprising the keyword query representation, the CJN representation, and the similarity score between them.

 Following our approach in QM ranking, we use a Bayesian model, as detailed in previous work \cite{Martins@ICDE23PyLatheDB,Martins@IEEAccess23Lathe,Oliveira@ICDE15CNRank}, to compute the relevance score of each CJN. This score forms the basis for our fine-tuning training set, allowing the model to learn patterns unique to CJN relevance and improve the accuracy of our transformer-based ranking.
A score of 1 indicates high relevance to the keyword query, while other scores are computed using a sigmoid function of the Bayesian model score, with an added weight of 0.4 to ensure negative examples are appropriately accounted for. The similarity score calculation is expressed as follows:

\[
    Sim(Q,S)= 
\begin{dcases}
    1, & \text{if } C\!J\!N \text{ is relevant for } Q\\
    \dfrac{1}{1 + e^{-0.4{\times}bayesian\_score(Q,C\!J\!N)}},              & \text{otherwise}
\end{dcases}
\]

where $S$ represents a sentence for the candidate joining network $C\!J\!N$, and $bayesian\_score(Q,C\!J\!N)$ denotes the Bayesian model score for query $Q$ in relation to $C\!J\!N$.

\begin{example}
	\label{ex:cjn-finetuning}
	For illustration, consider the sentences for the candidate joining networks $C\!J\!N_1$ and $C\!J\!N_2$, shown in Figure~\ref{fig:encoding-cjns}. From these, we generate the training examples shown in Figure~\ref{fig:cjns-trainig-examples}.
 
    \begin{figure}[!htb]
        \centering
        \begin{adjustbox}{max width=\textwidth}	
            {\renewcommand{\arraystretch}{1.2}
            \begin{tabular}{|c|>{\raggedright}p{9cm}|>{\raggedright\arraybackslash}p{9cm}|}
            \hline
            \textbf{Aggregation Approach}          & \textbf{Positive Examples}      & \textbf{Negative Examples}                                            \\ \hline
            \multirow{3}{*}{\textbf{Mean}} & ("query: Will Smith films", "answer: person.name: Smith, Will | movie: films", score=1.0) & ("query: Will Smith films", "answer: person.name: Wills, Luke | movie: films | person.name: Smith, Tom", score=0.044)      \\ \hhline{~--}
            & ("query: Will Smith films", "answer: person.name: Smith, Will | movie: films", score=1.0) & ("query: Will Smith films", "answer: person.name: Hussing, Will | movie: films | person.name: Smith, Andrew", score=0.044) \\ \hhline{~--}
            & ("query: Will Smith films", "answer: person.name: Smith, Will | movie: films", score=1.0) & ("query: Will Smith films", "answer: person.name: Woods, Will | movie: films | person.name: Smith, Cedric", score=0.044)   \\ \hline
            \textbf{Multivalue} &
              ("query: Will Smith films", "answer: person.name: Smith, Will, Smith, Will, Smith, Will | movie: films", score=1.0) &
              ("query: Will Smith films", "answer: person.name: Wills, Luke, Hussing, Will, Woods, Will | movie: films | person.name: Smith, Tom, Smith, Andrew, Smith, Cedric", score=0.044) \\ \hline
            \end{tabular}
            }
        \end{adjustbox}
        \caption{Training examples for the CJN ranking fine-tuning}
        \label{fig:cjns-trainig-examples}
    \end{figure}

\end{example}

By adapting sentence-transformer models specifically for CJN ranking, we improve the model's ability to assign higher relevance scores to CJNs that align closely with the keyword query. This adaptation ensures that the model captures complex patterns unique to relational data, ultimately enhancing the relevance and accuracy of search results.

\section{Data Augmentation}\label{sec:data-agumentation}

In order to facilitate the fine-tuning process, which necessitates a robust set of training examples comprising keyword queries, their respective relevant QMs, and CJNs, we employed a data augmentation strategy. Given the importance of having a diverse and comprehensive dataset for effective model training, data augmentation plays a pivotal role in enriching the available training examples.  One of the challenges in this context is the lack of annotated data for training. Thus, one of our goals is to minimize the need for manually labeling data by leveraging data augmentation techniques to generate a wider variety of training examples.

The data augmentation process involve four main steps: (i) Extraction of CJN Templates; (ii) Generation of new keyword queries and CJNs; (iii) Run the keyword search for each query; (iv) Generate sentences for each QM or CJN generated by {\metodo} for that query.

Given a keyword query and its relevant CJN, we initiate the augmentation process by extracting CJN templates. A template denotes a CJN wherein all keywords are replaced by wildcards, represented by the symbol `?'. This step facilitates the creation of generalized structures that encapsulate the essence of query semantics without being bound to specific keywords.

\begin{example}
	\label{ex:template-results}
	Considering the candidate joining networks $C\!J\!N_1$, which was generated in Example~\ref{ex:cjn-encoding} and it is the relevant CJN for the query ``Will Smith films''. We can extract the following template $T_1$, from $C\!J\!N_1$ by replacing its keywords with a wildcard `?'. This template selects information on all the movies for all persons, and its results are shown in Figure~\ref{fig:template-results}.

    \begin{tabular}{cc}    
        \tikz{
            \node[main node] (1) {$C\!J\!N_1=  PERSON^V[name^{ \{will,smith\} }]$};
            \node[main node] (2) [below = 3mm of 1] {$CASTING$};
            \node[main node] (3) [below = 3mm of 2] {$MOVIE^S[self^{\{films\}}]$};
            \path[draw,thick,->]
            (2) edge node {} (1)
            (2) edge node {} (3)
        }
        &
        \tikz{
            \node[main node] (1) {$T_1=  PERSON^V[name^{ \{?\} }]$};
            \node[main node] (2) [below = 3mm of 1] {$CASTING$};
            \node[main node] (3) [below = 3mm of 2] {$MOVIE^S[self^{\{?\}}]$};
            \path[draw,thick,->]
            (2) edge node {} (1)
            (2) edge node {} (3)
        }
    \end{tabular}
 
    \begin{figure}[!htb]
        \centering
        \begin{adjustbox}{max width=0.5\textwidth}	
            \begin{tabular}{|l|l|r|}
                \hline
                \multicolumn{3}{|l|}{
                    \textbf{Results for} $T_1$
                }
                \\ \hline 
                \textbf{t1.name} & \textbf{t3.title} & \textbf{t3.year} \\\hline
                Hues, Jack       & The Guardian          & 1990 \\\hline
                Coote, Robert    & The House of Fear     & 1939 \\\hline
                O'Halloran, Jack & The Flintstones       & 1994 \\\hline
                Zorn, John       & Notes on Marie Menken & 2006 \\\hline
                Kern, Robert     & Plymouth Adventure    & 1952\\
                \hline
            \end{tabular}
        \end{adjustbox}
        \caption{Results for the template $T_1$.}
        \label{fig:template-results}
    \end{figure}
\end{example}

Next, we run the SQL statement derived from $T_1$ against the database, which returns the rows shown in Figure~\ref{fig:template-results}. Then, for each row, we are able to generate a new keyword query, and its relevant CJN and QM. First, we use a subset of the row values to generate a keyword query. Second, we generate the relevant CJN by filling the wildcards with keywords from the query. Third, the relevant QM is the set of non-free keyword matches from the CJN. Figure~\ref{fig:synthetic-queries} shows five keyword queries and their answers, which were generated from template $T_1$.

\begin{figure}[!htb]
    \centering
    \begin{adjustbox}{max width=\textwidth}	
        {
        \renewcommand{\arraystretch}{1.5}
        \begin{tabular}{|c|l|l|}
        \hline
        \multicolumn{3}{|l|}{\textbf{Keyword Queries and Answers generated for template} $T_1$} \\ \hline
        \multirow{3}{*}{1} & Query & Hues Jack films     \\ 
        \hhline{~--}
                           & CJN           & \tikz[baseline]{
            \node[main node,anchor=base] (1) {$PERSON^V[name^{ \{hues,jack\} }]$};
            \node[main node] (2) [right = 3mm of 1] {$CASTING$};
            \node[main node] (3) [right = 3mm of 2] {$MOVIE^S[self^{\{films\}}]$};
            \path[draw,thick,->]
            (2) edge node {} (1)
            (2) edge node {} (3)
        }                    \\ 
        \hhline{~--}
                           & QM            & $\{PERSON^V(name^{\{hues,jack\}}), MOVIE^S(self^{\{films\}})\}$                    \\ \hline
        \multirow{3}{*}{2} & Query & Coot Robert films   \\ 
        \hhline{~--}
                           & CJN           &  \tikz[baseline]{
            \node[main node,anchor=base] (1) {$PERSON^V[name^{ \{coot,robert\} }]$};
            \node[main node] (2) [right = 3mm of 1] {$CASTING$};
            \node[main node] (3) [right = 3mm of 2] {$MOVIE^S[self^{\{films\}}]$};
            \path[draw,thick,->]
            (2) edge node {} (1)
            (2) edge node {} (3)
        }                   \\ 
        \hhline{~--}
                           & QM            & $\{PERSON^V(name^{\{coot,robert\}}), MOVIE^S(self^{\{films\}})\}$                    \\ \hline
        \multirow{3}{*}{3} & Query & Halloran Jack films \\ 
        \hhline{~--}
                           & CJN           & \tikz[baseline]{
            \node[main node,anchor=base] (1) {$PERSON^V[name^{ \{halloran,jack\} }]$};
            \node[main node] (2) [right = 3mm of 1] {$CASTING$};
            \node[main node] (3) [right = 3mm of 2] {$MOVIE^S[self^{\{films\}}]$};
            \path[draw,thick,->]
            (2) edge node {} (1)
            (2) edge node {} (3)
        }                    \\ 
        \hhline{~--}
                           & QM            &   $\{PERSON^V(name^{\{halloran,jack\}}), MOVIE^S(self^{\{films\}})\}$                   \\ \hline
        \multirow{3}{*}{4} & Query & Zorn John films     \\ 
        \hhline{~--}
                           & CJN           & \tikz[baseline]{
            \node[main node,anchor=base] (1) {$PERSON^V[name^{ \{zorn,john\} }]$};
            \node[main node] (2) [right = 3mm of 1] {$CASTING$};
            \node[main node] (3) [right = 3mm of 2] {$MOVIE^S[self^{\{films\}}]$};
            \path[draw,thick,->]
            (2) edge node {} (1)
            (2) edge node {} (3)
        }                    \\ 
        \hhline{~--}
                           & QM            &  $\{PERSON^V(name^{\{zorn,john\}}), MOVIE^S(self^{\{films\}})\}$                   \\ \hline
        \multirow{3}{*}{5} & Query & Kern Robert films   \\ 
        \hhline{~--}
                           & CJN           & \tikz[baseline]{
            \node[main node,anchor=base] (1) {$PERSON^V[name^{ \{kern,robert\} }]$};
            \node[main node] (2) [right = 3mm of 1] {$CASTING$};
            \node[main node] (3) [right = 3mm of 2] {$MOVIE^S[self^{\{films\}}]$};
            \path[draw,thick,->]
            (2) edge node {} (1)
            (2) edge node {} (3)
        }                    \\ 
        \hhline{~--}
                           & QM            & $\{PERSON^V(name^{\{kern,robert\}}), MOVIE^S(self^{\{films\}})\}$                    \\ \hline
        \end{tabular}
        }
    \end{adjustbox}
    \caption{Keyword Queries and and Answers for the template $T_1$.}
    \label{fig:synthetic-queries}
\end{figure}

We use these augmented data to generate positive examples for the fine-tuning of the QM ranking and CJN ranking. Next, we perform the keyword search for the newly created queries, and use the non-relevant QMs and CJNs returned by {\metodo} as negative examples.

\section{Experiments}\label{chap:experiments} 

In this section, we report a set of experiments performed using datasets and query sets previously used in similar experiments reported in the literature. Our goal is to evaluate the transformer-based QM Ranking and CJN Ranking proposed in this paper.

\subsection{Experimental Setup}
\label{subsec:setup}


\subsubsection*{Datasets and Query Sets}

\newcommand{\ftna}{\footnote{https://www.imdb.com/}}
\newcommand{\ftnb}{\footnote{https://www.cia.gov/library/publications/the-world-factbook/}}
\newcommand{\fttd}{\footnote{https://towardsdatascience.com/understanding-boxplots-5e2df7bcbd51}}
\newcommand{\ftnc}{\footnote{https://www.yelp.com/dataset}}

For all the experiments, we used three datasets, \emph{IMDb}, \emph{MONDIAL}, and \emph{Yelp}, which were used in the evaluation of previous R-KwS systems and methods \cite{Coffman@CIKM10Framework,Coffman@TKDE12Evaluation,Luo@SIGMOD07Spark,Oliveira@ICDE15CNRank,Oliveira@ICDE18MatCNGen,Oliveira@TKDE20,afonso@SBBD21SEREIA}. The IMDb dataset is a subset of the well-known Internet Movie Database (IMDb)\ftna, containing information related to films, television shows, and home videos – including actors, characters, and more. The MONDIAL dataset \cite{May@99MONDIAL} comprises geographical and demographic information from sources like the \emph{CIA World Factbook}\ftnb, the International Atlas, the TERRA database, and other web sources. The Yelp dataset is a subset of {Yelp}\ftnc, containing information about businesses, reviews, and user data.

The three datasets have distinct characteristics. The IMDb dataset has a simple schema, but query keywords often occur in several relations. Although the MONDIAL dataset is smaller, its schema is more complex or dense, with more relations and relational integrity constraints (RICs). The Yelp dataset has the highest number of tuples but its schema is simple. Table~\ref{tab:datasets_querysets} summarizes the details of each dataset and its respective query sets.

\begin{table}[htb]
	\centering
	\caption{Datasets we used in our experiments} 
	\label{tab:datasets_querysets}
		\small
\begin{tabular}{lrrrrrr} 
	\toprule
	Dataset & Size(MB) & Relations &  Attributes & RIC & Tuples & Total Queries    \\ 
	\midrule
    IMDb       &  701  &   6  &  33  &   5  &  1,673,076 & 50 \\
	MONDIAL    &   14  &  28  &  48  &  38  &  17,115    & 50 \\
    Yelp  &  7898  &   7  &  24  &   5  &  12,856,448   & 28 \\
	\bottomrule
\end{tabular} 
\end{table}

\subsubsection*{Golden Standards}

    The benchmark by Coffman \& Weaver~\cite{Coffman@CIKM10Framework} provided the relevant interpretation and its relevant SQL results for each query of the IMDb and MONDIAL datasets. In the case of the Yelp dataset, SQLizer~\cite{yaghmazadeh@ACMPL17SQLIZER} provided the relevant SQL queries from natural language queries. Since we derived keyword queries from the latter, we also adapted the SQL queries to reflect this change.
    We then manually generated the golden standards for CJNs and QMs using relevant SQL provided by Coffman \& Weaver and in SQLizer. These golden standards are available at \url{https://github.com/pr3martins/Lathe}.

\subsubsection*{Metrics}

We evaluate the ranking of CJNs and QMs using several metrics: Recall, Recall at ranking position $K$ ($R@K$), Max Recall Position, and Mean Reciprocal Rank~(MRR).

Recall is the ratio of relevant results retrieved to the total number of relevant results. Recall at $K$ ($R@K$) is the mean recall across multiple queries considering only the first $K$ results. If fewer than $K$ results are retrieved by a system, we calculate the recall value at the last result. For instance, if the system returns the relevant CJN in at most position $3$ of the ranking for 35 out of 50 queries, then the system would obtain an $R@3$ of $0.7$.

Given that each keyword query has exactly one relevant QM and one relevant CJN, $R@K$ effectively measures whether the relevant QM or CJN is among the top K results returned by the system. This metric is important for understanding the effectiveness of our retrieval approach in ensuring that the relevant results are generated and returned.

The Max Recall Position indicates the highest rank position $K$ within which the relevant QM is found. A lower Max Recall Position implies that the relevant QM is found within the top $K$ positions, reducing the need to generate CJNs for many QMs.

The Mean Reciprocal Rank (MRR) value indicates how close the correct CJN is to the first position of the ranking. Given a keyword query $Q$, the value of the \emph{reciprocal ranking for $Q$} is given by $RR_Q = \frac{1}{K}$, where $K$ is the rank position of the relevant result. Then, the MRR obtained for the queries in a query set is the average of $RR_Q$, for all $Q$ in the query set.
This metric is particularly useful in our context because it penalizes systems that place the relevant QM or CJN further down the ranking.  It, thus, highlights models that not only retrieve the relevant item but do so with higher precision.

\subsubsection*{{\metodo} Setup} 
For the experiments we report here, we set a maximum size for QMs and CJNs of 3 and 5, respectively. Also, we consider three important parameters for running {\metodo}: $N_{QM}$, the maximum number of QMs considered from the QM ranking; $N_{CJN}$, the maximum number of CJNs considered from each QM; 
and $P_{CJN}$, the number of CJNs probed per QM by the eager evaluation. In this context, a \emph{setup} for {\metodo} is a triple $N_{QM}/N_{CJN}/P_{CJN}$. The most common setup we used in our experiments is
$8/1/9$, in which we take the top-8 QMs in the ranking, generate and probe up to 9 CJNs for each QM, 
and take only the first non-empty CJN, if any, from each QM. We call this the \emph{default setup}.
Later in this section, we will discuss how these parameters affect the effectiveness and the performance 
of {\metodo}, as well as why we use the default configuration.

\subsubsection*{Fine-Tuning Setup} 

In our experiments, we used a set of parameters for configuring the transformer-based models we used 
as well as the data augmentation techniques we described in Section~\ref{sec:data-agumentation}. We utilized a batch size of 128, which was chosen as it was the maximum that did not exceed the GPU memory capacity, ensuring efficient utilization of hardware resources without causing memory overflow. The model was trained over 2 epochs, allowing us to go over the dataset twice to balance between sufficient training time and computational efficiency. We incorporated 100 warmup steps to gradually increase the learning rate, aiding in stable model convergence. Evaluations were conducted every 500 steps to monitor performance and make necessary adjustments during training.
We used the CosineSimilarityLoss function for training, ensuring that the model effectively learned to measure similarity between embeddings. This choice of loss function was crucial for our task, as it directly influenced the quality and accuracy of the learned embeddings.

For fine-tuning, we relied on the golden standards as the validation set, ensuring accurate performance evaluation against a trusted reference. The training and test sets were derived from data augmentation, with a generation ratio of 200. We split the data into 80\% for training and 20\% for testing, maintaining a balanced approach to model training and evaluation.

Additionally, we utilized view snapshots, comprising the first 3 rows from the database views, to generate sentences for the CJNs. This technique provided structured sentences that encapsulated essential information from the views, facilitating meaningful comparisons and relevance assessments.

\subsubsection*{System Details}

We ran the experiments reported on a Linux machine Ubuntu 22.04.4 LTS (64-bit, 32GB RAM, Intel(R) Xeon(R) W-2225 CPU @ 4.10GHz, 2 x 16GiB DIMM DDR4 Synchronous 3200 MHz, NVIDIA Quadro RTX 5000). We used PostgreSQL as the underlying RDBMS with a default configuration. All implementations were made in Python 3.

\subsection{Neural QM Ranking}\label{sec:neural-qmranking}

We evaluate the performance of different models on the task of QM ranking using the metrics MRR (Mean Reciprocal Rank), Recall, and Max Recall Position. Our goal is to assess the effectiveness of transformer-based models for ranking QMs and benchmark them against the Bayesian model, which serves as the ranking solution in our Bayesian approach.

A lower Max Recall Position ($k$) indicates that the relevant QM is found within the top $k$ positions, requiring the CJN generation only for the first $k$ QMs. Nevertheless, a high Max Recall Position generates a major amount of CJNs, this ensures that the relevant QM is generated for most keyword queries, thereby increasing the likelihood of generating the most relevant CJN. Hence, a high MRR indicates good overall ranking performance. 

Table~\ref{tab:neural_qm_ranking_models_short} shows twenty models analyzed for this experiment, detailing their similarity function and acronym. The letters $B$, $P$ and $F$ in the acronym respectively identify bayesian, pre-trained and fine-tuned models. The subscript indicator identifies each model, and in the case of transformer-based models it also identifies their respective variations. The Bayesian model ($B_{Q\!M}$) follows the Bayesian approach decribed in~\cite{Martins@IEEAccess23Lathe,Oliveira@ICDE15CNRank} and serves as a baseline for comparison against the transformer-based models. The analysis includes both pre-trained and fine-tuned models.

\begin{table}[!htb]
    \centering
    
    \caption{Neural QM Ranking Models}
    \label{tab:neural_qm_ranking_models_short}
    
        {
        \renewcommand{\arraystretch}{1.1}
        \begin{tabular}{|l|c|cc|}
\hline
\textbf{Model}                        & \textbf{Similarity} & \multicolumn{2}{c|}{\textbf{Abbreviation}}                      \\ \hline
Bayesian                                 & cos                 & \multicolumn{2}{c|}{$B_{QM}$}                                \\ \hline
\multirow{2}{*}{\textbf{Model}} & \multirow{2}{*}{\textbf{Similarity}} & \multicolumn{2}{c|}{\textbf{Abbreviation}} \\ 
\hhline{~~--}
                                      &                     & \multicolumn{1}{c|}{\textbf{Pre-Trained}} & \textbf{Fine-tuned} \\ \hline
paraphrase-albert-small-v2            & cos                 & \multicolumn{1}{c|}{$P_{Albert}$}         & $F_{Albert}$        \\ \hline
all-distilroberta-v1                  & cos                 & \multicolumn{1}{c|}{$P_{Distil1}$}        & $F_{Distil1}$       \\ \hline
distiluse-base-multilingual-cased-v1  & cos                 & \multicolumn{1}{c|}{$P_{Distil2}$}        & $F_{Distil2}$       \\ \hline
distiluse-base-multilingual-cased-v1  & dot                 & \multicolumn{1}{c|}{$P_{Distil3}$}        & $F_{Distil3}$       \\ \hline
distiluse-base-multilingual-cased-v2  & cos                 & \multicolumn{1}{c|}{$P_{Distil4}$}        & $F_{Distil4}$       \\ \hline
distiluse-base-multilingual-cased-v2  & dot                 & \multicolumn{1}{c|}{$P_{Distil5}$}        & $F_{Distil5}$       \\ \hline
multi-qa-distilbert-cos-v1            & cos                 & \multicolumn{1}{c|}{$P_{Distil6}$}        & $F_{Distil6}$       \\ \hline
all-MiniLM-L12-v2                     & cos                 & \multicolumn{1}{c|}{$P_{MiniLM1}$}        & $F_{MiniLM1}$       \\ \hline
all-MiniLM-L6-v2                      & cos                 & \multicolumn{1}{c|}{$P_{MiniLM2}$}        & $F_{MiniLM2}$       \\ \hline
all-MiniLM-L6-v2                      & dot                 & \multicolumn{1}{c|}{$P_{MiniLM3}$}        & $F_{MiniLM3}$       \\ \hline
multi-qa-MiniLM-L6-cos-v1             & cos                 & \multicolumn{1}{c|}{$P_{MiniLM4}$}        & $F_{MiniLM4}$       \\ \hline
multi-qa-MiniLM-L6-cos-v1             & dot                 & \multicolumn{1}{c|}{$P_{MiniLM5}$}        & $F_{MiniLM5}$       \\ \hline
paraphrase-MiniLM-L3-v2               & cos                 & \multicolumn{1}{c|}{$P_{MiniLM6}$}        & $F_{MiniLM6}$       \\ \hline
paraphrase-multilingual-MiniLM-L12-v2 & dot                 & \multicolumn{1}{c|}{$P_{MiniLM7}$}        & $F_{MiniLM7}$       \\ \hline
paraphrase-multilingual-MiniLM-L12-v2 & cos                 & \multicolumn{1}{c|}{$P_{MiniLM8}$}        & $F_{MiniLM8}$       \\ \hline
all-mpnet-base-v2                     & cos                 & \multicolumn{1}{c|}{$P_{MPNET1}$}         & $F_{MPNET1}$        \\ \hline
all-mpnet-base-v2                     & dot                 & \multicolumn{1}{c|}{$P_{MPNET2}$}         & $F_{MPNET2}$        \\ \hline
multi-qa-mpnet-base-dot-v1            & dot                 & \multicolumn{1}{c|}{$P_{MPNET3}$}         & $F_{MPNET3}$        \\ \hline
paraphrase-multilingual-mpnet-base-v2 & cos                 & \multicolumn{1}{c|}{$P_{MPNET4}$}         & $F_{MPNET4}$        \\ \hline
\end{tabular}%
        }
\end{table}


\subsubsection*{All Datasets Analysis}

\begin{figure*}[!htb]
     \centering
     \adjustbox{width=\textwidth}{
        \begin{tabular}{cc}
             \includegraphics[width=\textwidth]{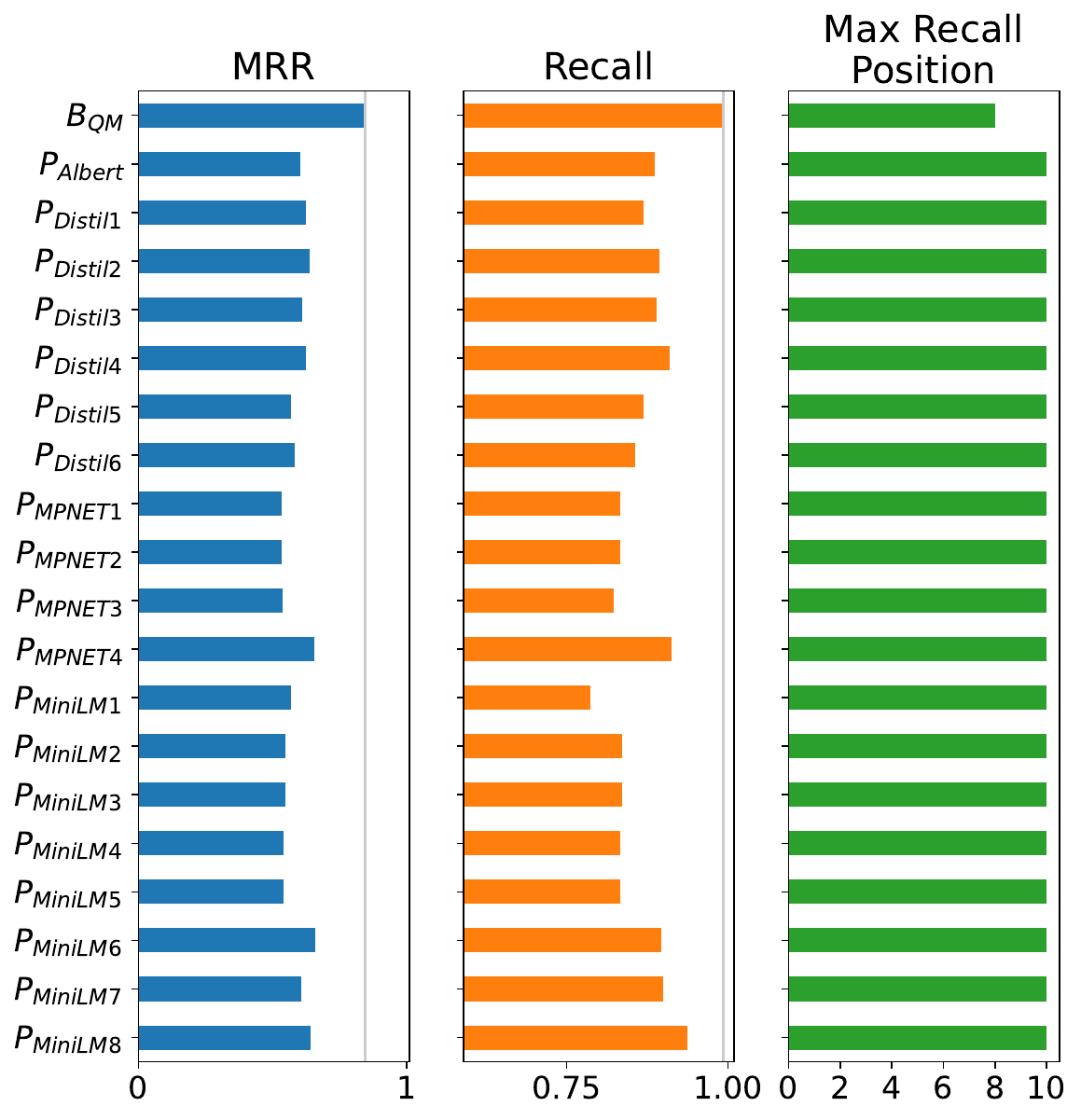}
             &
             \includegraphics[width=\textwidth]{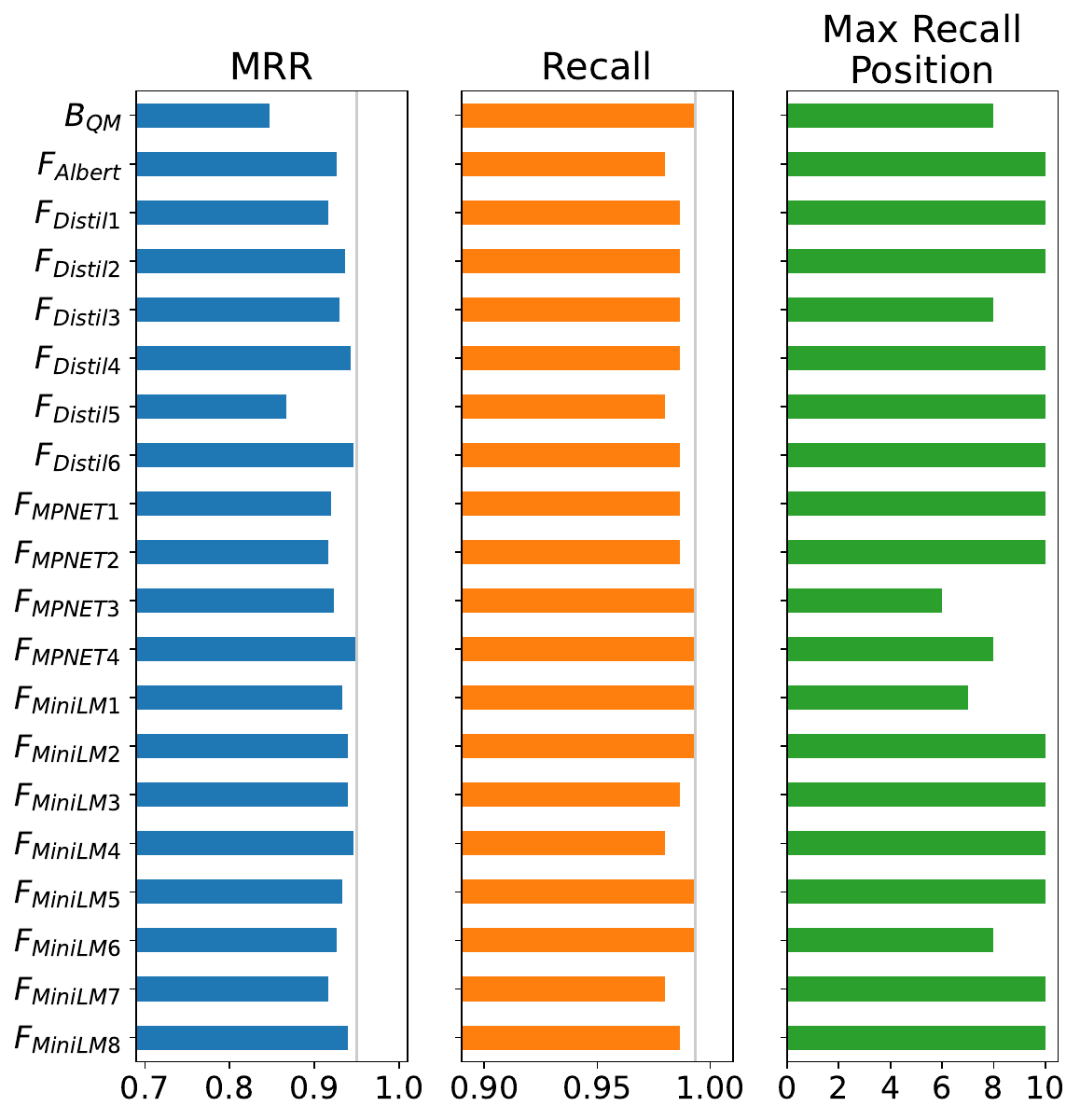}
             \\
             {\LARGE (a) Pre-trained Models}
             &
             {\LARGE (b) Fine-tuned Models}
        \end{tabular}
    }
    \caption{Evaluation of the Neural QM ranking on all datasets (Average for MMR and Recall, and Max for Max Recall Position). 
    }
    \label{plot:neural_qm_alldatasets}
\end{figure*}

Figure~\ref{plot:neural_qm_alldatasets} shows the average MRR and Recall across the IMDB, MONDIAL, and Yelp datasets, along with the maximum Max Recall Position for the same datasets.

The Bayesian model ($B_{Q\!M}$) achieved an MRR of 0.847, Recall of 0.993, and a Max Recall Position of 8. Pre-trained models  performed worse than the Bayesian model, which indicates they struggled to accurately interpret QM linearization and ranking tasks. However, fine-tuning these models led to substantial improvements. 

On the other hand, fine-tuned models like $F_{MPNET4}$, $F_{MiniLM4}$, and $F_{Distil6}$ achieved higher MRR scores than the baseline, indicating their superior capability in ranking relevant QMs higher on average. This highlights the impact of fine-tuning in enhancing the models' understanding and ranking abilities for QM tasks.

Regarding Recall, several fine-tuned models, including $F_{MPNET3}$, $F_{MPNET4}$, $F_{MiniLM1}$, $F_{MiniLM2}$, $F_{MiniLM5}$, and $F_{MiniLM6}$, achieved a Recall of 0.993. This performance is on par with the Bayesian baseline, showcasing these models' ability to retrieve all relevant QMs effectively. Such results underscore the models' comprehensive recall capabilities after fine-tuning.

In terms of Max Recall Position, the fine-tuned models also demonstrated improvements. For instance, $F_{MPNET3}$ achieved the best performance by retrieving the relevant QM within the top 6 positions, outperforming the baseline's position 8. Similarly, $F_{MiniLM1}$ and $F_{MPNET4}$ achieved positions 7 and 8, respectively. These results reflect the models' enhanced efficiency in identifying relevant QMs earlier in the ranking list, a crucial factor for user satisfaction in information retrieval tasks.

Among the evaluated models, $F_{MiniLM1}$, $F_{MPNET3}$, and $F_{MPNET4}$ stand out as the best performers overall. These models consistently achieve high MRR, Recall, and low Max Recall Position across all datasets.

\subsubsection*{Individual Dataset Analysis}

The results for individual datasets largely mirror the trends observed in the overall analysis, with fine-tuned models outperforming pre-trained ones and the Bayesian baseline. 

\begin{figure*}[!htb]
     \centering
     \adjustbox{width=\textwidth}{
        \begin{tabular}{cc}
             \includegraphics[width=\textwidth]{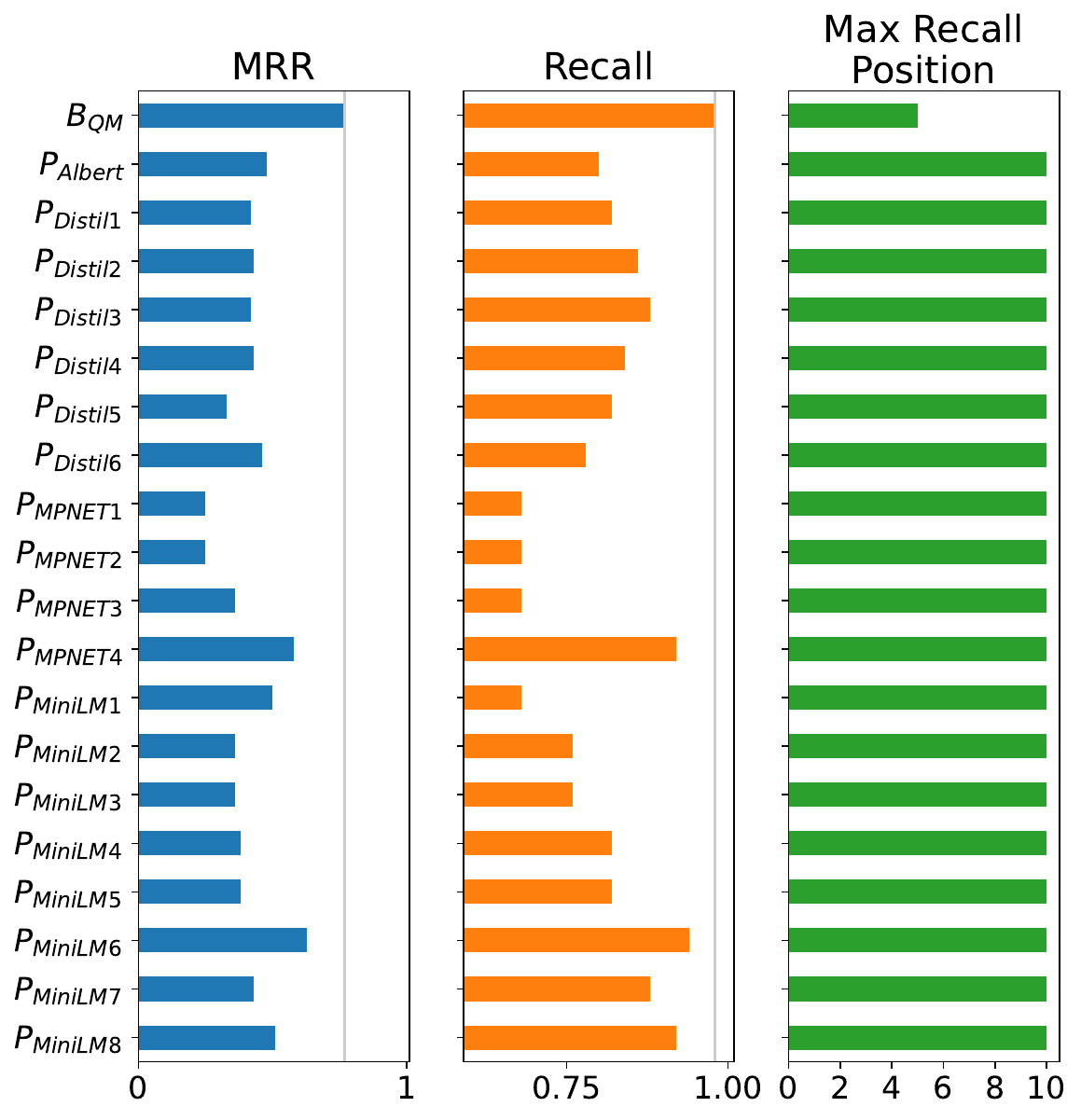}
             &
             \includegraphics[width=\textwidth]{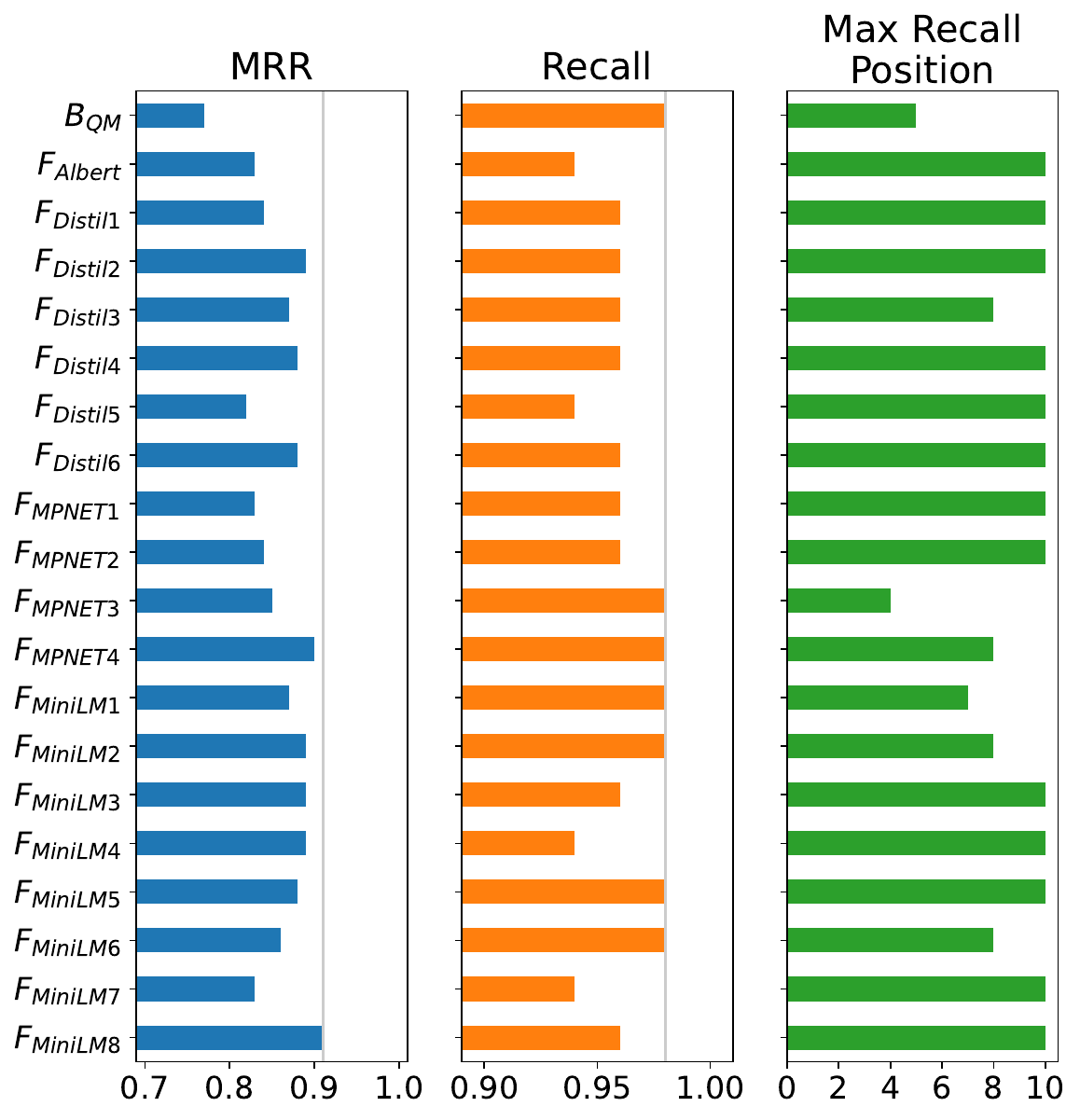}
             \\
             {\LARGE (a) Pre-trained Models}
             &
             {\LARGE (b) Fine-tuned Models}
        \end{tabular}
    }
    \caption{Evaluation of the Neural QM ranking on the IMDb dataset.
    }
    \label{plot:neural_qm_imdb}
\end{figure*}

Figure~\ref{plot:neural_qm_imdb} shows the results for the IMDB dataset. The Bayesian model ($B_{Q\!M}$) achieved an MRR of 0.77, Recall of 0.98, and a Max Recall Position of 5. Pre-trained models consistently performed worse than the baseline. Fine-tuned models like $F_{MiniLM8}$ and $F_{MPNET4}$ achieved MRR scores of 0.91 and 0.9, respectively, indicating their superior capability in ranking relevant QMs higher on average. 
Several fine-tuned models, including $F_{MPNET3}$, $F_{MPNET4}$, $F_{MiniLM1}$, $F_{MiniLM2}$, $F_{MiniLM5}$, and $F_{MiniLM6}$, achieved a Recall of 0.98, on par with the Bayesian baseline, showcasing these models' ability to retrieve all relevant QMs effectively.  In terms of Max Recall Position, $F_{MPNET3}$ achieved the best performance by retrieving the relevant QM within the top 4 positions, outperforming the baseline's position 5. Similarly, $F_{MiniLM1}$ and $F_{MPNET4}$ achieved positions 7 and 8, respectively, reflecting enhanced efficiency in identifying relevant QMs earlier.

\begin{figure*}[!htb]
     \centering
     \adjustbox{width=\textwidth}{
        \begin{tabular}{cc}
             \includegraphics[width=\textwidth]{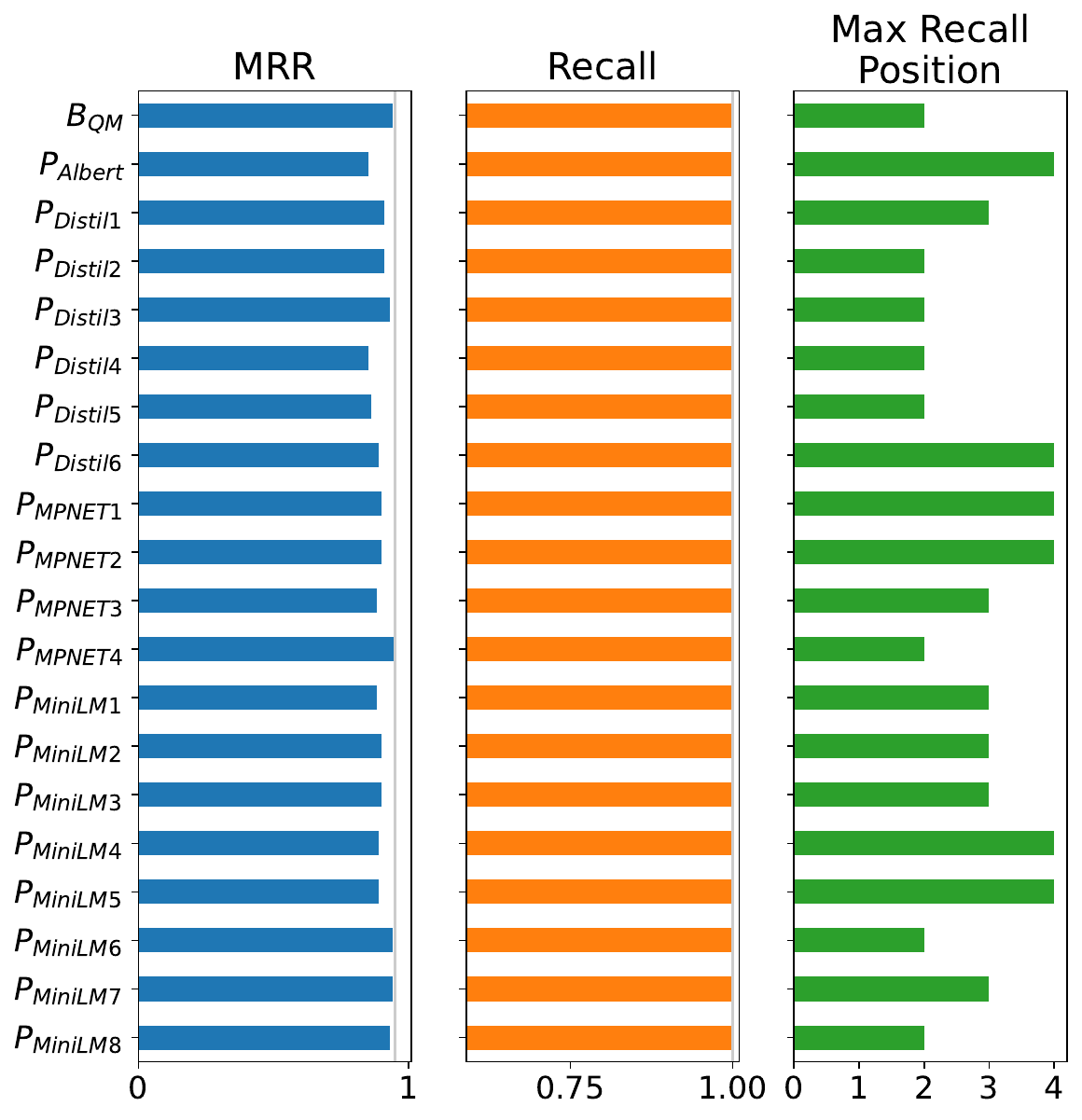}
             &
             \includegraphics[width=\textwidth]{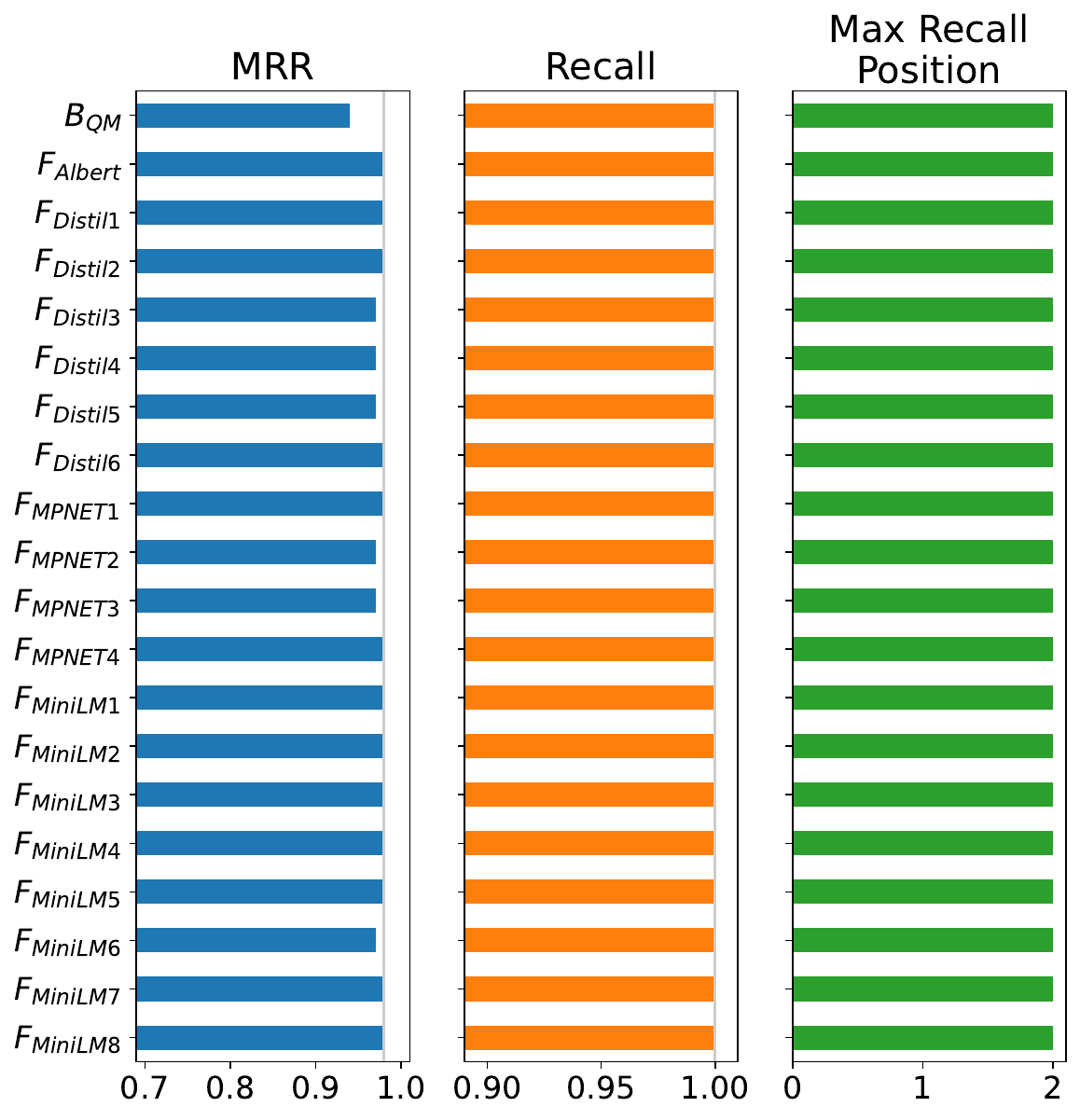}
             \\
             {\LARGE (a) Pre-trained Models}
             &
             {\LARGE (b) Fine-tuned Models}
        \end{tabular}
    }
    \caption{Evaluation of the Neural QM ranking on the MONDIAL dataset. 
    }
    \label{plot:neural_qm_mondial}
\end{figure*}

Figure~\ref{plot:neural_qm_mondial} shows the results for the MONDIAL dataset. The Bayesian model ($B_{Q\!M}$) achieved an MRR of 0.94, Recall of 1.0, and a Max Recall Position of 2. Most sentence-transformer models achieved perfect Recall (1.0) and very low Max Recall Positions (2). The pre-trained model $P_{MPNET4}$ obtained a slightly better MRR of 0.95. Fine-tuned models like $F_{Albert}$, $F_{Distil1}$, $F_{MPNET4}$, and $F_{MiniLM1}$ all achieved an MRR of 0.98, making them top performers for this dataset. Due to MONDIAL having fewer QMs on average, 
a pre-trained model achieved a better MRR score than the baseline, and models based on the Albert and the distilled version of RoBERTa were among the top performers.

\begin{figure*}[!htb]
     \centering
     \adjustbox{width=\textwidth}{
        \begin{tabular}{cc}
             \includegraphics[width=\textwidth]{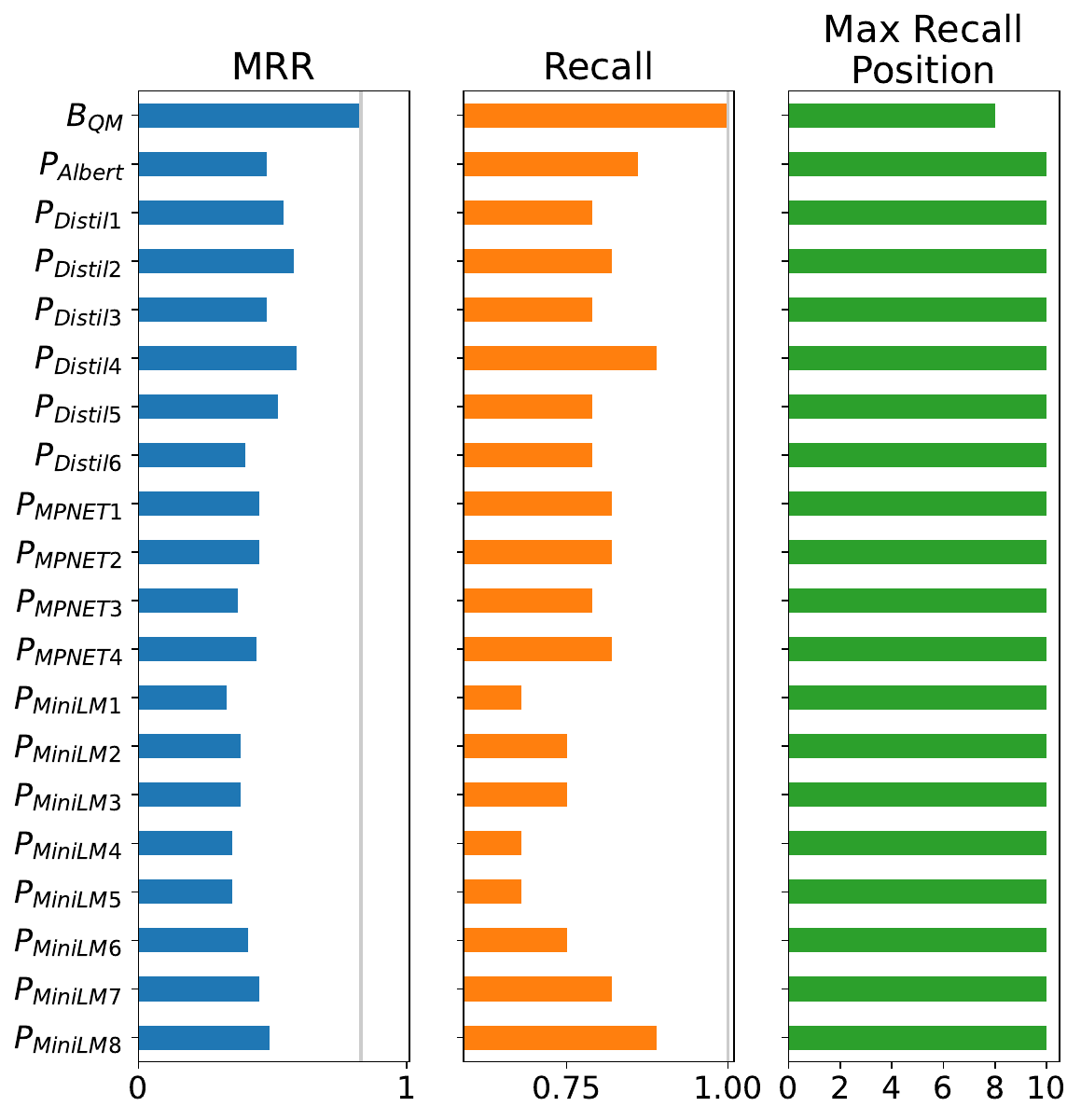}
             &
             \includegraphics[width=\textwidth]{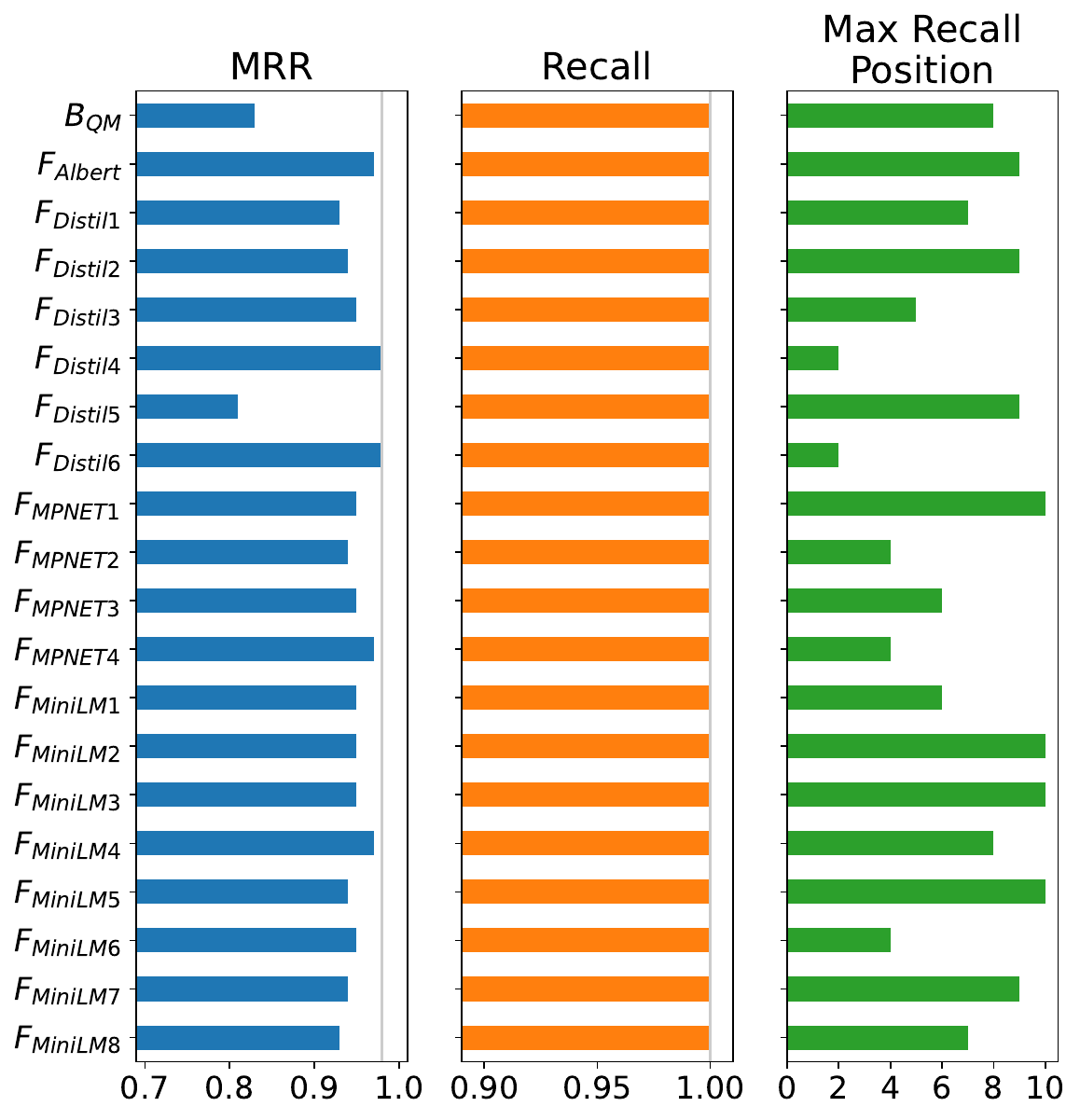}
             \\
             {\LARGE (a) Pre-trained Models}
             &
             {\LARGE (b) Fine-tuned Models}
        \end{tabular}
    }
    \caption{Evaluation of the Neural QM ranking on the Yelp dataset. 
    }
    \label{plot:neural_qm_yelp}
\end{figure*}

Figure~\ref{plot:neural_qm_yelp} shows the results for the Yelp dataset. The Bayesian model ($B_{Q\!M}$) had an MRR of 0.83, Recall of 1.0, and a Max Recall Position of 8. Pre-trained models consistently performed worse than the baseline. Fine-tuned models like $F_{Distil4}$ and $F_{Distil6}$ achieved the highest MRR of 0.98 and Recall of 1.0, with a Max Recall Position of 2. However, $F_{MPNET4}$ and $F_{MiniLM1}$ also performed well, obtaining MRR scores of 0.97 and 0.95 respectively, and Max Recall Positions of 4 and 6.

\subsubsection*{Discussion}

The results indicate that sentence-transformer models significantly outperform the Bayesian baseline in terms of MRR, suggesting that these models provide better overall ranking of QMs. Fine-tuned models, in particular, demonstrate superior performance in the QM ranking.

Models based on MiniLM and MPNET were consistently among the top performers across all datasets, a trend also observed in the  experiments shown on the Sentence-Transformers library website\footnote{Sentence Transformers Pretrained Models https://sbert.net/docs/sentence\_transformer/pretrained\_models.html}. Specifically, the models $F_{MiniLM1}$, $F_{MPNET3}$, and $F_{MPNET4}$ achieved high MRR, Recall, and low Max Recall Position consistently, making them ideal candidates for practical applications where efficiency in QM ranking is crucial. Consequently, these models were chosen as the best performers for the QM ranking task and were utilized in the Neural CJN ranking experiment.

\subsection{Neural CJN Ranking} \label{sec:neural-cjnranking}


Table~\ref{tab:neural_cjn_ranking_models} shows twenty-four models analyzed for this experiment, detailing their similarity function and abbreviation. The letters $B$, $P$, and $F$ in the abbreviations respectively identify Bayesian, pre-trained, and fine-tuned models. The subscript indicator identifies each model, and in the case of transformer-based models, it also identifies their respective variations. Each abbreviation consists of its QM model abbreviation followed by a dash, and then the CJN abbreviation. 
The CJN abbreviation may also have trailing characters indicating the aggregation approach used, which can be `*' for the multivalue approach, and `+' for the mean approach.
The CJN abbreviation $B_{C\!J\!N}$ indicates that a simple CJN ranking was applied, where the CJNs were ranked based on their QM score, divided by their size.
The model $B_{Q\!M}{-}B_{C\!J\!N}$ follows the Bayesian approach, and it serves as a baseline for comparison against the transformer-based models. 
These models were ranked in the top-10 in at least one of the datasets. 

\begin{table}[!htb]
    \centering
    \caption{CJN Ranking Models}
    \tiny
    \renewcommand{\arraystretch}{1.5}
        \begin{tabular}{|l|l|c|c|c|l|}
\hline
\textbf{QM Model} & \textbf{CJN Model}                    & \textbf{Similarity} & \textbf{Agg. Approach} & \textbf{CJN Type} & \textbf{Abbreviation}          \\ \hline
$B_{Q\!M}$   & Bayesian                                & cos &    ---        & bayesian   & $B_{Q\!M}{-}B_{C\!J\!N}$       \\ \hline
$B_{Q\!M}$   & paraphrase-albert-small-v2            & cos & Multivalue & fine-tuned & $B_{Q\!M}{-}F_{Albert{*}}$    \\ \hline
$B_{Q\!M}$   & all-distilroberta-v1                  & cos & Multivalue & fine-tuned & $B_{Q\!M}{-}F_{Distil1{*}}$   \\ \hline
$B_{Q\!M}$   & multi-qa-MiniLM-L6-cos-v1             & dot & Mean       & fine-tuned & $B_{Q\!M}{-}F_{MiniLM5{+}}$   \\ \hline
$B_{Q\!M}$   & paraphrase-multilingual-MiniLM-L12-v2 & cos & Multivalue & fine-tuned & $B_{Q\!M}{-}F_{MiniLM8{*}}$   \\ \hline
$F_{MiniLM1}$ & Bayesian                                & cos & ---     & bayesian   & $F_{MiniLM1}{-}B_{C\!J\!N}$     \\ \hline
$F_{MiniLM1}$ & paraphrase-albert-small-v2            & cos & Multivalue & fine-tuned & $F_{MiniLM1}{-}F_{Albert{*}}$  \\ \hline
$F_{MiniLM1}$ & all-distilroberta-v1                  & cos & Multivalue & fine-tuned & $F_{MiniLM1}{-}F_{Distil1{*}}$ \\ \hline
$F_{MiniLM1}$ & all-distilroberta-v1                  & cos & Mean       & fine-tuned & $F_{MiniLM1}{-}F_{Distil1{+}}$ \\ \hline
$F_{MiniLM1}$     & distiluse-base-multilingual-cased-v1  & dot                 & Multivalue             & fine-tuned        & $F_{MiniLM1}{-}F_{Distil3{*}}$ \\ \hline
$F_{MiniLM1}$ & all-MiniLM-L6-v2                      & cos & Multivalue & fine-tuned & $F_{MiniLM1}{-}F_{MiniLM2{*}}$ \\ \hline
$F_{MiniLM1}$ & all-MiniLM-L6-v2                      & dot & Multivalue & fine-tuned & $F_{MiniLM1}{-}F_{MiniLM3{*}}$ \\ \hline
$F_{MiniLM1}$ & paraphrase-MiniLM-L3-v2               & cos & Mean       & fine-tuned & $F_{MiniLM1}{-}F_{MiniLM6{+}}$ \\ \hline
$F_{MiniLM1}$     & paraphrase-multilingual-MiniLM-L12-v2 & cos                 & Multivalue             & fine-tuned        & $F_{MiniLM1}{-}F_{MiniLM8{*}}$ \\ \hline
$F_{MPNET3}$  & Bayesian                                & cos & ---     & bayesian   & $F_{MPNET3}{-}B_{C\!J\!N}$      \\ \hline
$F_{MPNET3}$  & paraphrase-albert-small-v2            & cos & Multivalue & fine-tuned & $F_{MPNET3}{-}F_{Albert{*}}$   \\ \hline
$F_{MPNET3}$  & all-distilroberta-v1                  & cos & Multivalue & fine-tuned & $F_{MPNET3}{-}F_{Distil1{*}}$  \\ \hline
$F_{MPNET3}$  & distiluse-base-multilingual-cased-v1  & dot & Multivalue & fine-tuned & $F_{MPNET3}{-}F_{Distil3{*}}$  \\ \hline
$F_{MPNET3}$  & all-MiniLM-L6-v2                      & dot & Multivalue & fine-tuned & $F_{MPNET3}{-}F_{MiniLM3{*}}$  \\ \hline
$F_{MPNET3}$  & paraphrase-MiniLM-L3-v2               & cos & Mean       & fine-tuned & $F_{MPNET3}{-}F_{MiniLM6{+}}$  \\ \hline
$F_{MPNET3}$  & paraphrase-multilingual-MiniLM-L12-v2 & cos & Multivalue & fine-tuned & $F_{MPNET3}{-}F_{MiniLM8{*}}$  \\ \hline
$F_{MPNET4}$  & Bayesian                                & cos & ---     & bayesian   & $F_{MPNET4}{-}B_{C\!J\!N}$      \\ \hline
$F_{MPNET4}$  & distiluse-base-multilingual-cased-v1  & dot & Multivalue & fine-tuned & $F_{MPNET4}{-}F_{Distil3{*}}$  \\ \hline
$F_{MPNET4}$  & paraphrase-multilingual-MiniLM-L12-v2 & cos & Multivalue & fine-tuned & $F_{MPNET4}{-}F_{MiniLM8{*}}$  \\ \hline
\end{tabular}%
    \renewcommand{\arraystretch}{1}
    \label{tab:neural_cjn_ranking_models}
\end{table}

\subsubsection*{Evaluation on All Datasets}

\begin{figure*}[!htb]
	\centering
	\includegraphics[width=\linewidth]{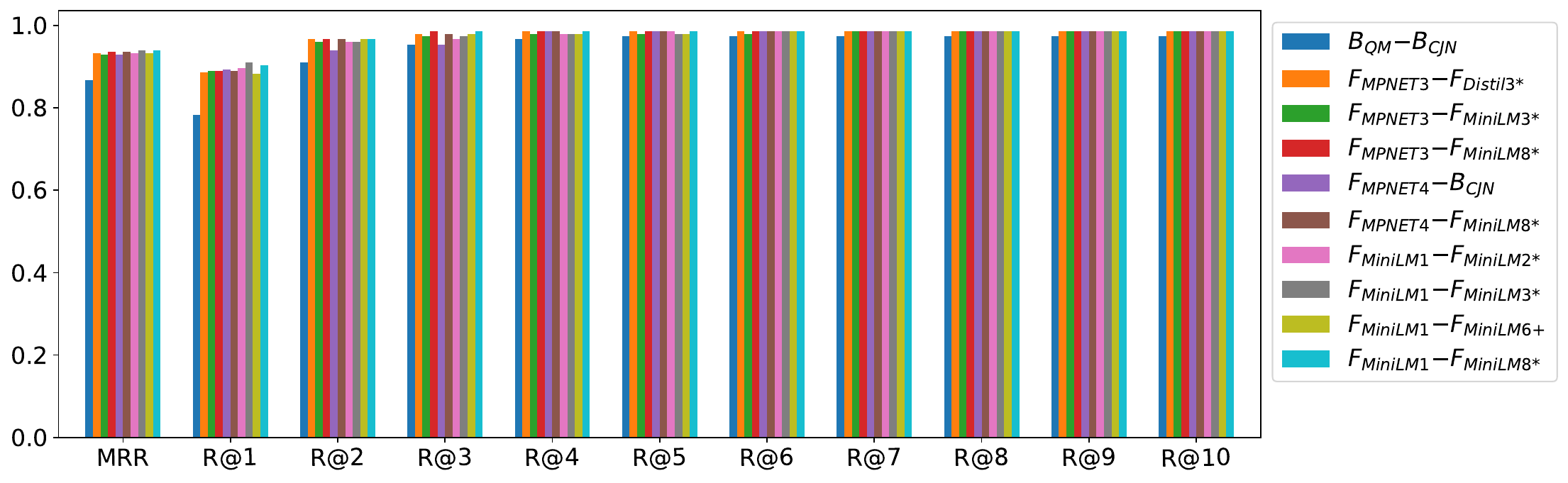}
	\caption{Neural CJN Ranking - All Datasets}
	\label{plot:cjn-ranking-allqueryset}
\end{figure*}

We begin by evaluating the performance of various CJN models on all available query sets. Figure~\ref{plot:cjn-ranking-allqueryset} presents the Mean Reciprocal Rank (MRR) and Recall@k (R@k) scores for each model.

The baseline model, $B_{Q\!M}{-}B_{C\!J\!N}$, achieves a decent performance with an MRR of 0.867 and high Recall scores, with R@1 at 0.7833, R@2 at 0.91, reaching a recall plateau of 0.9733 at R@6.
However, most of the fine-tuned models show substantial improvements over the baseline, with MRR values ranging from 0.93 to 0.94. Recall metrics for these models also indicate enhanced performance, often obtaining a R@1 above 0.883, R@2 above 0.94, and reaching or exceeding 0.98 by R@4.

Specifically, the models $F_{MPNET3}{-}F_{Distil3{*}}$, $F_{MPNET3}{-}F_{MiniLM8{*}}$, and $F_{MiniLM1}{-}F_{MiniLM8{*}}$ are the top performers, with MRR values of 0.9333, 0.9367, and 0.94, respectively. These models demonstrate superior recall performance compared to the baseline. For instance, $F_{MPNET3}{-}F_{Distil3{*}}$ achieves an R@1 of 0.8867 and an R@10 of 0.9867, showing consistent improvement over the baseline across all recall points.

Models involving fine-tuning on MiniLM and MPNET consistently outperform others, which aligns with trends observed in the pretraining experiments on the Sentence-Transformers library website\footnote{Sentence Transformers Pretrained Models https://sbert.net/docs/sentence\_transformer/pretrained\_models.html}.

Among the top 10 models, one uses the mean aggregation approach, seven use the multivalue aggregation approach, and one model besides the baseline employs the simple ranking method. This distribution highlights the superiority of the multivalue aggregation approach for generating embeddings for CJNs. Additionally, the inclusion of the $F_{MPNET4}{-}B_{C\!J\!N}$ model among the top performers suggests that the $F_{MPNET4}$ model is so effective in QM ranking that there is no need to rank CJNs using another neural model. Instead, we can simply divide the QM score by the size of the CJN to achieve competitive results.

Furthermore, the combination of models, such as $F_{MPNET3}{-}F_{MiniLM8{*}}$, indicates that pairing different fine-tuned models can leverage their strengths, leading to improved ranking accuracy.

Overall, the fine-tuned models, especially those based on MiniLM and MPNET, exhibit substantial improvements in MRR and recall metrics, demonstrating their effectiveness in the CJN ranking task. The consistent high performance across various recall points underscores their potential for practical applications in information retrieval systems.

\subsubsection*{Evaluation on Specific Datasets}

We further evaluate the CJN models on specific datasets to assess their performance in domain-specific scenarios. The results for individual datasets largely mirror the trends observed in the overall analysis, with fine-tuned models outperforming pre-trained ones and the Bayesian baseline. 

\begin{figure*}[!htb]
	\centering
	\includegraphics[width=\linewidth]{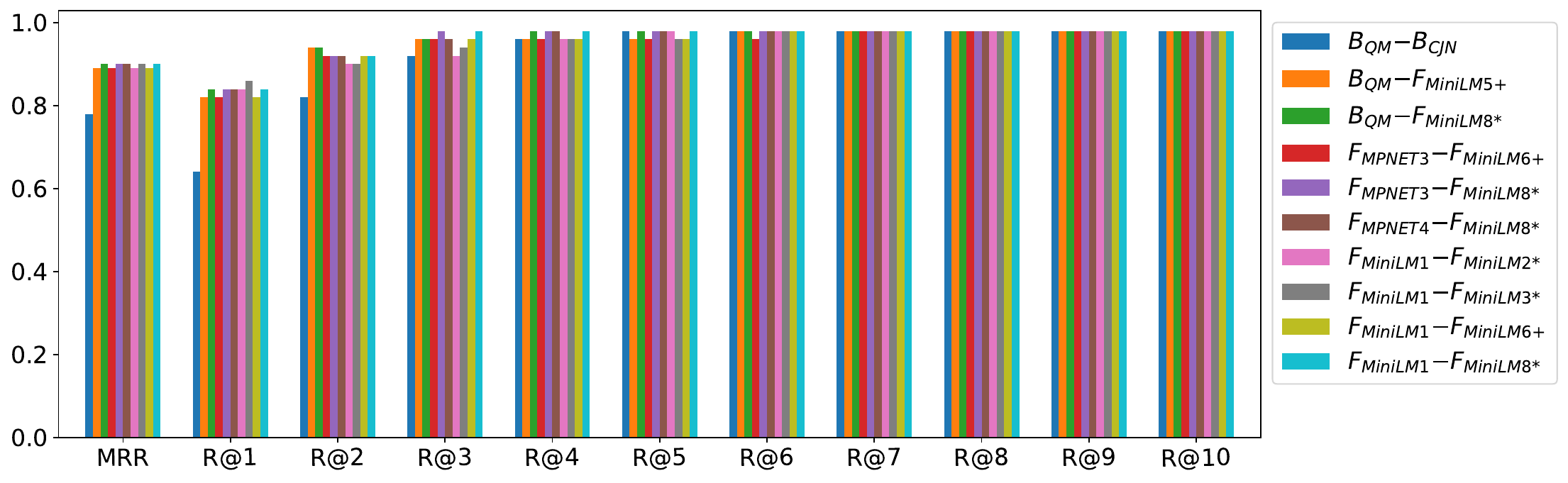}
	\caption{Neural CJN Ranking - IMDb}
	\label{plot:cjn-ranking-imdb}
\end{figure*}

Figure~\ref{plot:cjn-ranking-imdb} shows the results for the IMDb dataset.  
The baseline model $B_{Q\!M}{-}B_{C\!J\!N}$ achieved an MRR of 0.78, highlighting a significant improvement with transformer-based models.
Fine-tuned models demonstrate better performance at lower recall points (R@1 to R@4), indicating their effectiveness in ranking relevant CJNs higher. Most models achieve a high recall plateau at R@6.
The fine-tuned models $B_{Q\!M}{-}F_{MiniLM8{*}}$, $F_{MPNET3}{-}F_{MiniLM8{*}}$, $F_{MPNET4}{-}F_{MiniLM8{*}}$, $F_{MiniLM1}{-}F_{MiniLM8{*}}$ achieved the highest MRR of 0.9 and consistently high recall scores. This indicates that the $F_{MiniLM8{*}}$ model is highly effective for the CJN ranking, no independent of which model was used for the QM ranking. Furthermore, all the top 4 performer models used the multivalue (*) approach, indicating its superiority.

\begin{figure*}[!htb]
	\centering
	\includegraphics[width=\linewidth]{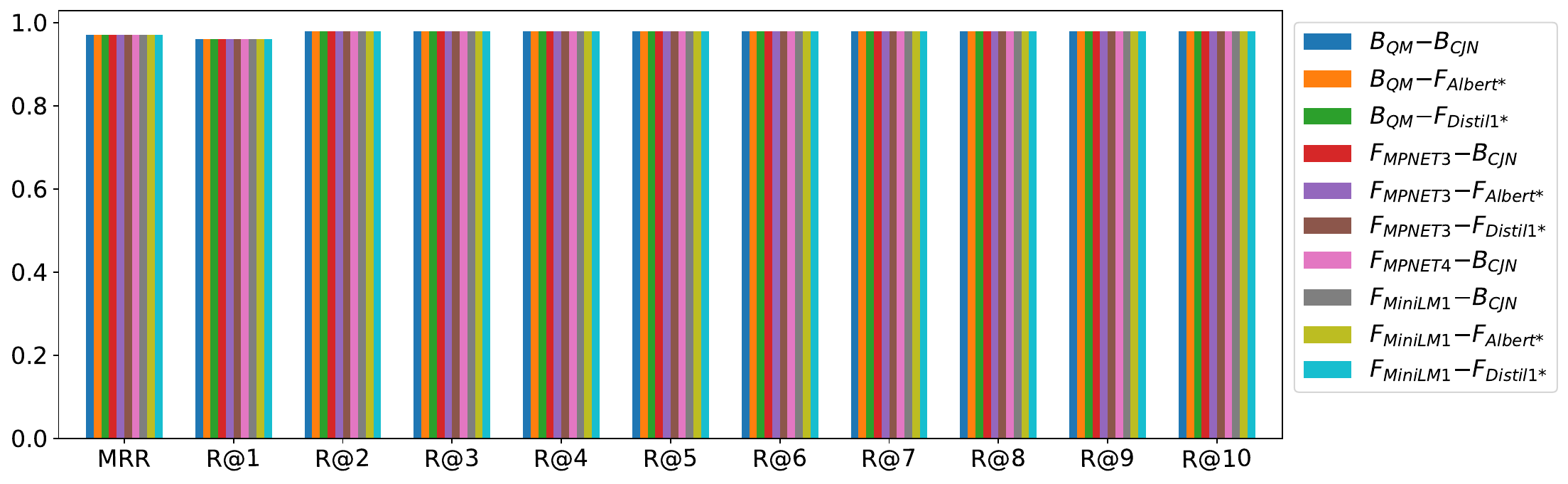}
	\caption{Neural CJN Ranking - MONDIAL}
	\label{plot:cjn-ranking-mondial}
\end{figure*}

Figure~\ref{plot:cjn-ranking-mondial} shows the results for the MONDIAL dataset. The baseline model $B_{Q\!M}{-}B_{C\!J\!N}$ performs equally well compared to fine-tuned models. This performance equality might be attributed to the dataset's nature, where simpler models can achieve high performance due to less complexity in the data.
All models achieved the same MRR of 0.97, and reached a recall plateau at R@2, indicating that there is minimal variation in performance across different CJN models for this dataset.

\begin{figure*}[!htb]
	\centering
	\includegraphics[width=\linewidth]{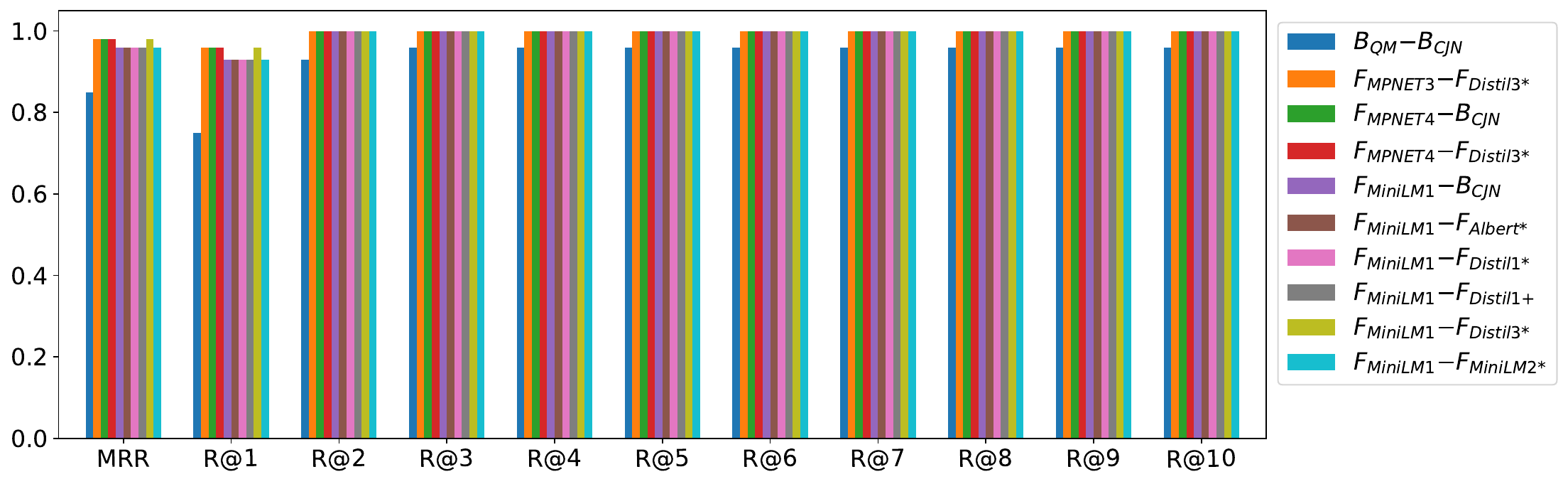}
	\caption{Neural CJN Ranking - Yelp}
	\label{plot:cjn-ranking-yelp}
\end{figure*}

Figure~\ref{plot:cjn-ranking-yelp} shows the results for the Yelp dataset. The baseline model $B_{Q\!M}{-}B_{C\!J\!N}$ had an MRR of 0.85, significantly lower than the top-performing fine-tuned models, highlighting the efficacy of transformer-based approaches. Several fine-tuned models achieve perfect recall from R@2 onwards, indicating these models are highly effective at ranking the most relevant CJNs at the top. Models like $F_{MPNET3}{-}F_{Distil3{*}}$, $F_{MPNET4}{-}F_{Distil3{*}}$, and $F_{MiniLM1}{-}F_{Distil3{*}}$ achieved an MRR of 0.98, with perfect recall from R@2 onwards, demonstrating outstanding performance in ranking relevant CJNs.

\subsubsection*{Discussion}

The comprehensive evaluation of CJN models across different datasets highlights the superiority of fine-tuned transformer models over traditional Bayesian models. In the IMDB and Yelp datasets, fine-tuned models demonstrated substantial improvements in Mean Reciprocal Rank (MRR) and recall scores, particularly at lower recall points. This underscores the ability of transformer models to effectively rank the most relevant CJNs at the top positions, making them highly suitable for practical applications in information retrieval systems.

In the IMDB dataset, models like $B_{Q\!M}{-}F_{MiniLM8{*}}$ and $F_{MPNET3}{-}F_{MiniLM8{*}}$ achieved the highest MRR of 0.9, far surpassing the baseline's 0.78. The consistent high performance across various recall points further demonstrates their reliability in identifying relevant context join networks. The use of multivalue aggregation approaches played a crucial role in this success.

For the Yelp dataset, fine-tuned models reached near-perfect or perfect recall scores, with models such as $F_{MPNET3}{-}F_{Distil3{*}}$ achieving an MRR of 0.98. These models consistently ranked the most relevant CJNs at the top positions, showcasing the effectiveness of transformer-based approaches over traditional Bayesian methods for high-accuracy relevance ranking.

In contrast, the Mondial dataset revealed that both Bayesian and fine-tuned transformer models performed equally well, with all models achieving an MRR of 0.97. This suggests that for datasets with less complexity, simpler models can be as effective as more complex transformer-based approaches. The minimal performance variation across different CJN models for this dataset highlights the suitability of simpler models in certain contexts, where the added complexity of fine-tuning may not yield significant advantages.

Overall, the findings indicate that fine-tuned transformer models, particularly those involving MiniLM and MPNET, offer substantial improvements in MRR and recall metrics across various datasets. The consistent high performance underscores their potential for enhancing information retrieval systems. However, the effectiveness of simpler models in certain datasets suggests that the choice of model should be context-dependent, balancing complexity and performance based on the dataset characteristics.

\subsection{Final Remarks}

In the experiments reported, we compared the Bayesian and Neural Ranking approaches for CJN and QM ranking tasks. The Bayesian ranking method is a fully unsupervised approach, simplifying its implementation and deployment. On the other hand, the Neural Ranking approach has shown significant improvements in metrics such as MRR and R@K for CJN ranking and has also enhanced the QM ranking, although it required fine-tuning to achieve these results.

Performance-wise, although we omit detailed results here for brevity, additional experiments revealed that the Bayesian approach consistently runs faster for both QM ranking and CJN ranking when compared to transformer-based models. Despite this speed advantage, neural models outperformed the Bayesian method in key effectiveness metrics such as MRR and recall. With the ongoing advancement of neural architectures and computational hardware, we anticipate that transformer-based models will become increasingly practical and may eventually surpass the Bayesian approach in both speed and quality.

\section{Conclusion}\label{chap:conclusion}

This paper tackled the challenges of accurately ranking Query Matches (QMs) and Candidate Joining Networks (CJNs) in Relational Keyword Search (R-KwS) systems. Building on Lathe —- a state-of-the-art method for generating CJNs capable of handling both schema and attribute-value references -- we proposed a new transformer-based ranking approach.

Our main contributions can be summarized as follows:

We introduced novel transformer-based strategies for ranking QMs and CJNs that leverage sentence-transformer models to generate richer, context-aware embeddings for both QMs and CJNs. For this, we propose a linearization process that converts these structures into textual sequences, allowing us to effectively apply transformer models to these tasks.

We developed a fine-tuning strategy that adapts pre-trained transformer models to the relational keyword search context. This process not only enhanced the models’ ability to rank QMs and CJNs accurately but also allowed them to generalize across different relational schemas and query types, contributing to their robustness in real-world scenarios. To address the scarcity of labeled training examples for fine-tuning, we devised a data augmentation pipeline that automatically generates a wide array of synthetic queries and their corresponding relevant CJNs and QMs. 


We conducted an in-depth evaluation using well-known public datasets  and their associated query workloads. Our results demonstrate that transformer-based models significantly improve ranking quality. Notably, the fine-tuned \textit{MiniLM} and \textit{MPNET}-based models consistently outperformed the baseline across datasets, achieving higher MRR, recall, and lower Max Recall Position. Moreover, we showed that employing a \emph{multivalue} snapshot aggregation of CJN results can further boost ranking performance.

\subsection*{Future Work}

The proposed transformer-based ranking approach for Relational Keyword Search (R-KwS) demonstrated significant improvements in semantic relevance and ranking quality. However, there are several avenues for further exploration and enhancement.

Improving the efficiency of transformer-based models remains a priority. Future efforts could explore techniques such as knowledge distillation, model pruning, and quantization to reduce resource consumption without compromising accuracy. Adopting advanced transformer architectures such as ModernBERT~\cite{Warner24} or lightweight variants may also further improve efficiency.

Developing methods to dynamically adapt the ranking process based on user feedback during query execution could enhance the system's relevance. Real-time learning mechanisms that leverage historical query data and user preferences would enable personalized and iterative improvements in query ranking.

Exploring the use of prompt engineering with advanced language models (e.g., GPT-4, Llama, or Mistral) represents another promising avenue. This approach could enhance the system's ability to interpret and rank Query Matches (QMs) and Candidate Joining Networks (CJNs) by better capturing the semantic and contextual nuances of user queries.

Introducing active learning frameworks could reduce the reliance on large annotated datasets. By selectively querying users to label ambiguous QMs or CJNs, the system could achieve incremental and cost-effective improvements. Continuous learning mechanisms could further refine model performance over time without requiring complete retraining. 

\section*{Acknowledgments}

This work was supported by the National Council for Scientific and Technological Development – CNPq, with a postdoc grant to Paulo Martins (Proc. 151880/2024-7),  individual grants to Altigran da Silva (Proc. 307248/2019-4), Edleno Moura (Proc. 310573/2023-8) and Project IAIA (Proc. 406417/2022-9); by FAPEAM under the POSGRAD 2022 Program and a posdoc grant to Johny Moreira in the context of the NeuralBond Project (UNIVERSAL 2023 Proc. 01.02.016301.04300/2023-04); and by the Coordination for the Improvement of Higher Education Personnel-Brazil (CAPES) financial code 001.

\bibliography{sn-bibliography}

\end{document}